\documentclass[a4paper,11pt]{article}
\pdfoutput=1 

\usepackage{jheppub} 
\usepackage{hyperref}
\usepackage[T1]{fontenc} 
\usepackage[normalem]{ulem}

\usepackage{color}

\usepackage{graphicx}
\usepackage{caption}
\usepackage{subcaption}
\usepackage{dcolumn}
\usepackage{bm}

\usepackage{color,xcolor,url,ulem,verbatim,appendix}
\usepackage[colorlinks=true
,urlcolor=blue
,anchorcolor=blue
,citecolor=blue 
,filecolor=blue
,linkcolor=blue
,menucolor=blue
,linktocpage=true
,pdfproducer=medialab
,pdfa=true
]{hyperref}
\usepackage{lineno}
\usepackage{bm,color,xcolor}
\usepackage{slashed} 
\usepackage{graphicx}
\usepackage{soul} 
\usepackage{multirow}
\usepackage{comment}
\usepackage{epstopdf} 
\usepackage{mathtools}
\usepackage{appendix}
\usepackage[colorlinks=true
,urlcolor=blue
,anchorcolor=blue
,citecolor=blue 
,filecolor=blue
,linkcolor=blue
,menucolor=blue
,linktocpage=true
,pdfproducer=medialab
,pdfa=true
]{hyperref}

\newcommand{\beq}{\begin{equation}}
\newcommand{\eeq}{\end{equation}}
\newcommand{\bea}{\begin{eqnarray}}
\newcommand{\eea}{\end{eqnarray}}
\newcommand{\ba}{\begin{array}}
\newcommand{\ea}{\end{array}}

\def\m1{M_1}
\def\m2{M_2}
\def\m3{M_3}

\def\ch10{\tilde \chi^0_1}

\def\TeV{\,{\rm TeV}}
\def\gev{\,{\rm GeV}}

\newcommand{\lsim}{\mathrel{\mathop{\kern 0pt \rlap
  {\raise.2ex\hbox{$<$}}}
  \lower.9ex\hbox{\kern-.190em $\sim$}}}
\newcommand{\gsim}{\mathrel{\mathop{\kern 0pt \rlap
  {\raise.2ex\hbox{$>$}}}
  \lower.9ex\hbox{\kern-.190em $\sim$}}}
\begin{document}
\title{ \bf \Large 
Sleptonic SUSY: From UV Framework to IR Phenomenology
}

\author[a]{Kaustubh Agashe,}
\emailAdd{kagashe@umd.edu}
\author[b]{Majid Ekhterachian,}
\emailAdd{majid.ekhterachian@epfl.ch}
\author[c]{Zhen Liu,} 
\emailAdd{zliuphys@umn.edu}
\author[a]{Raman Sundrum}
\emailAdd{raman@umd.edu}
\affiliation[a]{Maryland Center for Fundamental Physics, Department of Physics,
University of Maryland, College Park, MD 20742, USA}
\affiliation[b]{Theoretical Particle Physics Laboratory (LPTP),
Institute of Physics, EPFL, Lausanne, Switzerland}
\affiliation[c]{School of Physics and Astronomy, University of Minnesota, Minneapolis, MN 55455, USA}

%
%

\date{\normalsize  \today}

\abstract{ 
We study an attractive scenario, ``Sleptonic SUSY'', which reconciles the $125$ GeV Higgs scalar and   
the non-observation of superpartners thus far with 
potentially pivotal roles for slepton phenomenology: 
providing viable ongoing targets for LHC discovery, incorporating
a co-annihilation partner for detectable thermal relic dark matter, 
and capable of mediating  the potential muon $g-2$ anomaly.
This is accomplished by a modestly hierarchical spectrum, with sub-TeV sleptons and 
electroweakinos and with multi-TeV masses for the other new states. 
We study new elements in the UV MSSM realization of Sleptonic SUSY based on higher-dimensional sequestering and the 
synergy between the resulting gaugino-mediation, hypercharge $D$-term mediation and Higgs-mediation of SUSY-breaking, so as to more fully capture the range of possibilities. This framework stands out by harmoniously solving the flavor, CP and
 $\mu - B\mu$ problems of the supersymmetric paradigm. We discuss its extension to orbifold GUTs, including gauge-coupling and 
  $b$-tau unification. 
  We also develop a non-minimal model with extra Higgs fields, in which the electroweak vacuum is more readily cosmologically stable against decay to a charge-breaking vacuum, allowing a broader range of sleptonic spectra than in the MSSM alone. 
We survey the rich set of signals possible at the LHC and future colliders, covering both $R$-parity conservation and violation, as well as for dark matter detection.
While the multi-TeV squarks imply a Little Hierarchy Problem, intriguingly, small changes in parameter space to improve naturalness result in dramatic phase transitions to either electroweak-preservation  or charge-breaking. 
In a Multiverse setting, the modest unnaturalness may then be explained by the ``principle of living dangerously''. 
}

\preprint{UMD-PP-022-03
}

\maketitle


\section{Introduction}
Supersymmetry (SUSY) offers a very attractive and comprehensive framework for particle physics beyond the Standard Model (BSM) (for a review, see \cite{Martin:1997ns}). It is a subtle spacetime symmetry that is consistent with perturbatively renormalizable interactions, and is a very plausible ``remnant'' of superstring dynamics of quantum gravity. It is capable of beautifully addressing the electroweak (EW) hierarchy problem, improving the fit to Grand Unification, and offering WIMP dark matter (DM) candidates, 
if SUSY breaking occurs close to the weak scale. In general, as in any rich weak-scale BSM scenario, it faces the challenge of understanding how excessive flavor-changing neutral currents (FCNCs) and CP violation are suppressed as well as internal challenges such as the $\mu$ problem, but there are now robust
field theory mechanisms known, such as gauge-mediated SUSY breaking (GMSB) (for a review, see, for example, \cite{ Giudice:1998bp}) and 
the higher-dimensional sequestered structures of
anomaly-mediated SUSY breaking AMSB \cite{Randall:1998uk} and 
gaugino-mediated SUSY breaking (${\rm \tilde{g}}$MSB) \cite{Chacko:1999mi, Kaplan:1999ac}.

Despite these attractions, superpartners have not yet been seen at the LHC (or  at lower-energies) and, indirectly, the 125 GeV Higgs mass is most straightforwardly
accounted for by having stops at or above $~ 10$ TeV 
(for a review, see, for example,
\cite{Draper:2016pys}, 
%
%
where however the theory becomes significantly fine-tuned $\sim 10^{-3} - 10^{-4}$.\footnote{
It is well-known that one can lower the stop masses significantly in the MSSM while still accounting for the $125$ GeV Higgs mass, by 
including large SUSY-breaking $A$-terms (see for example, Ref. \cite{Draper:2016pys}). However, in the framework of gaugino-mediated supersymmetry breaking that we will be led to focus on in this paper, such terms do not  significantly improve the fine-tuning. Also, given existing search constraints, there is little room remaining for discovery at the LHC of such light  stops or gluinos. We will therefore not pursue the direction of large $A$-terms here.
}

This is the supersymmetric version of the general and puzzling clash between naturalness expectations and the absence of BSM physics to date, now known as the  ``Little Hierarchy Problem'' (LHP): see, for example, ref. \cite{Barbieri:2000gf}. 

The ``Split SUSY'' paradigm \cite{Wells:2003tf, Wells:2004di, Arkani-Hamed:2004ymt, Giudice:2004tc, Arkani-Hamed:2004zhs} introduced the possibility that SUSY is a remnant of UV structure, but {\it not} strongly tied to the EW hierarchy problem. Rather, the EW hierarchy problem is assumed to be solved by
 the Anthropic Principle 
 \cite{Carter:1974zz, Carr:1979sg,Barrow:1986nmg,Weinberg:1987dv,Vilenkin:1994ua,Agrawal:1997gf, Agrawal:1998xa}
operating within a very large multiverse \cite{Bousso:2000xa, Arkani-Hamed:2005zuc, Denef:2007pq}, in such a way that it undercuts the requirement of  SUSY naturalness.  Further, it considered the robust possibility that R-symmetry is only weakly broken in the SUSY breaking dynamics, so that the gauginos are naturally far lighter than the scalars. In this way, the scalars might lie far above LHC reach, while the gauginos could still be at the TeV scale and provide excellent WIMP DM candidates, while evading detection thus far.
In later ``Mini-Split SUSY'' variants \cite{Arvanitaki:2012ps, Arkani-Hamed:2012fhg}, a more modest hierarchy is considered between TeV-scale gauginos and 10 TeV scalars that more straightforwardly accounts for the observed 125 GeV Higgs (see, for example, \cite{Draper:2016pys}), with the viewpoint being that the resulting little hierarchy problem is resolved by 
some unspecified combination of naturalness and anthropic ``pressures''.

In this paper, we reconsider the big and little hierarchy problems in the context of  an attractive  scenario, which we will call ``Sleptonic SUSY'', that exhibits mechanisms to explain what we see and do  not see, and might still see, at low-energy experiments, the LHC and  future colliders, and dark matter detection experiments. Here too, there is a moderately hierarchical spectrum, but  with
colored superpartner masses $\sim {\cal O}(10)$ TeV,  while the sleptons and EW gauginos are sub-TeV. As in Mini-Split SUSY the ``heavy'' 125 GeV Higgs mass and multi-TeV squarks
go hand in hand.
But now there exists the interesting option of weak scale WIMP Dark Matter arising from slepton-bino co-annihilations \cite{Griest:1990kh,Drees:1992am,Ellis:1998kh} in the early universe, in contrast to the super-TeV Wino or Higgsino DM candidates of (Mini-) Split SUSY. While in (Mini-)Split SUSY, the classic DM candidates of winos or Higgsinos are very difficult to detect at the LHC (if they are thermal relics), here weak-scale sleptons as co-annihilation partners of DM may well be observable at the LHC, and the bino DM may be observable in direct detection experiments.

Sleptonic SUSY is further motivated by the potential anomaly in the muon $g-2$: the measurements \cite{Muong-2:2006rrc, Muong-2:2021ojo} seem to differ from the understanding of SM theory presented in \cite{Aoyama:2020ynm} at $\sim 4 \; \sigma$
level. 

See ref. \cite{Athron:2021iuf} for a review of SUSY (and non-SUSY) 
models addressing this anomaly, and 
refs. \cite{Cox:2021nbo,Chakraborti:2021dli,Baer:2021aax,Aboubrahim:2021xfi,Wang:2021bcx,Li:2021pnt,Ellis:2021zmg,Chakraborti:2021bmv,Endo:2021zal,Iwamoto:2021aaf,Baum:2021qzx,Frank:2021nkq,Heinemeyer:2021opc,Shafi:2021jcg,Gomez:2022qrb, Cao:2021tuh, Ibe:2021cvf, BhupalDev:2021ipu} for recent SUSY models that appeared after the Fermilab $g-2$ measurement
\cite{Muong-2:2021ojo}.  Combined with a suitable flavor-safe SUSY breaking mechanism, weak scale slepton exchange provides a simple way of getting an enhanced BSM muon $g-2$ contribution, while not conflicting with flavor/CP constraints. There are two notable enhancements that can take place and help explain the sheer size of the anomaly, from large $\tan \beta$ from the $2$-Higgs-doublet structure of the Minimal Supersymmetric Standard Model (MSSM) and from a multi-TeV $\mu$ term that can naturally accompany the large squark masses.  Such large $\mu$ then implies Higgsinos above a TeV.

While the Sleptonic SUSY spectrum is phenomenologically plausible and interesting, 
it is highly non-trivial to find a UV completion that encompasses mechanisms for flavor and CP safety, with a potential to address the muon $g-2$ discrepancy, and satisfies the current LHC constraints. See refs.~\cite{Harigaya:2015kfa, Yin:2016shg, Yanagida:2018eho, Bhattacharyya:2018inr, Yanagida:2020jzy, Cox:2018vsv, Ibe:2021cvf} for the recent literature.  
Along these lines, we consider ref.~\cite{Cox:2018vsv}, building on  ref.~\cite{Harigaya:2015kfa},
to be the most attractive model in the literature so far.
%
%
It is based on 
$\tilde{g}$MSB
%
%
%
\cite{Chacko:1999mi, Kaplan:1999ac}, which provides a flexible departure point for flavor-safe
 model-building to realize  the Sleptonic SUSY hierarchy, rooted in a hierarchical UV gaugino mass parameters.
The Sleptonic SUSY spectrum provides intriguing physics discovery opportunities at the LHC and future colliders. At the LHC, the weak-scale sleptons and electroweakinos are kinematically accessible, and one can perform new searches in this regime looking for  lepton-rich final states in which the lepton are not particularly hard, and doing combined analysis of multiple channels.
Further, multiple discoveries are guaranteed at future colliders, such as those of sleptons, electroweakinos as well as colored  superpartners, enabling us to definitively establish the SUSY character of the underlying physics.
 
Here, we make the case that multiple bottom-up phenomenological and top-down field-theoretic considerations combine to make Sleptonic SUSY a highly attractive BSM framework, deserving the strongest possible efforts to explore on the experimental front. We will bring together different separately-known mechanisms and show how they 
harmonize within the Sleptonic SUSY framework. While our perspective is somewhat different, 
our work includes generalization of the analysis and model of ref.\cite{Cox:2018vsv} to realize a greater variety of key phenomenological behaviors, including some of those  discussed 
from a more bottom-up perspective in follow-up work \cite{Cox:2021nbo}.

 In \autoref{sec:gauginomediation}, we will briefly review the departure point of ${\rm \tilde{g}}$MSB, emphasizing the elegant and plausible means by which it solves the thorny SUSY $\mu$, CP and flavor problems using higher-dimensional sequestering \cite{Randall:1998uk, Luty:1999cz, Kachru:2007xp}, making it a front-runner among  BSM paradigms.\footnote{
  Remarkably, the 4D UV boundary conditions of  ${\rm \tilde{g}}$MSB were anticipated in ref. \cite{Inoue:1991rk} 
 as an elegant (curved) superspace ansatz. However, from the purely 4D perspective this ansatz is somewhat puzzling since it is not radiatively stable, that is it can only hold at one energy scale. But in the higher-dimensional construction this ansatz is dynamically justified as the tree-level matching condition to the 4D effective field theory at the Kaluza-Klein scale.}
 Furthermore, we will see that it is straightforward to model non-universal gaugino masses in the UV which will be the prerequisite for realizing Sleptonic SUSY in the IR.
  In \autoref{sec:OneLoopRGE}, we provide the  1-loop Renormalization Group (RG) evolution of SUSY breaking from the UV. The gaugino-mediated effects are famously flavor-blind, being proportional to gauge couplings.  There
  are additional flavor-blind effects proportional to gauge couplings arising from hypercharge $D$-terms. The general parameter space of such effects was not explored in ref. \cite{Cox:2018vsv}, but we will see that it can have dramatic  phenomenological consequences, so it is worth pursuing here within a flavor/CP-safe UV framework. 
   There are also 
   ``Higgs-mediated'' effects
  \cite{Yamaguchi:2016oqz,Yin:2016shg, Yanagida:2018eho, Yanagida:2020jzy}
  proportional to Yukawa couplings, such as the well-known top-Yukawa-dependent running of Higgs mass. 
  But in addition, there are ordinarily subdominant effects, via the $\tau$ Yukawa coupling, which become significant with the hierarchical spectrum of Sleptonic SUSY, 
 combined with large $\tan \beta$.
  Even though Higgs-mediated effects are not flavor-blind, they do not lead to new flavor-violation because, in an extension of the GIM-mechanism, they are automatically diagonal in the fermion mass basis. In ref. \cite{Cox:2018vsv} the viable parameter space was found to lie at very large $\tan \beta \sim 50$, where the analysis is a complex mix of competing effects, while we find new regions of parameter space with lower $\tan \beta$ where the analysis is more flexible and transparent. Furthermore, there are parts of the new regions of the parameter space in 
    which bino DM is observable by direct detection experiments.
     But in the MSSM, in such a regime  the muon $g-2$ correction is much smaller than the current anomaly.

  In \autoref{sec:bottomup}, we motivate Sleptonic SUSY from an alternate perspective. We begin by 
    {\it not} assuming the SUSY paradigm, but rather considering  that  the muon $g-2$ anomaly is eventually fully validated by theory and experiment. We explore what kind of new physics could straightforwardly explain this sizeable effect while maintaining the GIM mechanism for simply explaining why new flavor/CP violating effects are absent in existing low-energy data. We argue from this point of view that Sleptonic SUSY enjoys qualitatively attractive features, compared to known alternatives, when paired with a flavor-safe SUSY-breaking mechanism such as gaugino-mediation. 
  In \autoref{sec:1loopconsiderations} we show 
 that at one-loop approximation it is straightforward to 
 understand all the qualitatively important features of 
  a viable Sleptonic SUSY spectrum within the MSSM, which also addresses the muon $g-2$ anomaly and co-annihilating DM.

While we argue that Sleptonic SUSY is an attractive fit to many of the phenomenological considerations and constraints on the SUSY solution to the greater hierarchy problem, the fact that it suffers from the 
LHP
%
%
%
(and that 
superpartners have not been explicitly discovered at the LHC) appears at first sight to undercut its plausibility. As in (Mini-)Split SUSY we can well imagine that the anthropic principle \cite{Weinberg:1987dv} operating in a large multiverse \cite{Bousso:2000xa} of competing effective field theories (EFT) might modify the principle of Naturalness. But understanding the physics and distribution of EFTs in the multiverse created by eternal inflation is highly challenging at this time, to say the least. The only robust corollary we can draw is the Principle of Living Dangerously \cite{Weinberg:1987dv, Arkani-Hamed:2005zuc}, 
%
%
whereby our universe may exhibit some degree of unnaturalness in its EFT if more natural portions of the parameter space would be clearly inhospitable to intelligent life. It is therefore interesting to check if Sleptonic SUSY, with its moderately unnatural little hierarchy problem, is ``living dangerously''. In \autoref{sec:livingdanerously}, we show that, remarkably, this is indeed the case. In particular, we show that for small changes of parameters that make the theory considerably more natural, sleptons condense and give the photon a mass, eliminating the central non-gravitational long-range force of Nature! This lends some credence to the anthropic principle lying at the root of the little hierarchy problem in this setting. We will refer this situation, where naturalness and anthropic criteria are in mild tension giving rise to a relatively mild apparent breakdown of perfect naturalness, as ``Frustrated Naturalness'' \cite{Ramantalk}.

In \autoref{sec:6D}, we discuss a straightforward higher-dimensional framework in which Orbifold GUTs (for a review, see for example \cite{Hall:2002ea} and references therein) and Sequestered SUSY-breaking are implemented in the UV \cite{Dermisek:2001hp,Haba:2002ve,Buchmuller:2005ma}.
We show how this can be done, consistent with the non-universal gaugino hierarchy needed to realize Sleptonic SUSY in the IR, while preserving gauge-coupling unification. Furthermore, we show how the heaviest generation (third) can satisfy $b-\tau$ Yukawa unification, while the light generations do not, based on their higher-dimensional ``geography''. Again, this section is a synthesis of pre-existing mechanisms in the literature.

While the one-loop RG evolution of couplings and SUSY-breaking effects from the GUT scale down to observable energies gives the qualitative features of Sleptonic SUSY in our framework, there are quantitatively important two-loop effects. In \autoref{sec:twoloopRGE}, we present the two-loop RG contributions that become important due to the hierarchical spectrum of Sleptonic SUSY. This completes the RG equations at the precision needed to derive the phenomenological consequences of the UV set-up. A simple observation at two-loop order is that the slepton-squark hierarchy cannot be arbitrarily large, unlike the hierarchy in Split SUSY between gauginos and scalars.
It is bounded by a loop factor at most, so that having weak scale sleptons requires squarks to be at most ${\cal O}(10)$ TeV in mass.

The issue of electroweak vacuum stability, already mentioned above in connection with the 
LHP, has two aspects. The simplest is the requirement that the correct electroweak VEVs at least represent a local minimum of the effective potential, in particular none of the sleptons should be tachyonic and thereby unstable to condensation. 
But even if this requirement is met, if the electroweak vacuum does not represent the absolute minimum of energy, then we must ensure that its lifetime is at least the age of our universe \cite{Rattazzi:1995gk,Rattazzi:1996fb,Hisano:2010re,Carena:2012mw,Kitahara:2013lfa}. This presents a more stringent requirement on the parameter space, and restriction of the phenomenology.  
We discuss meta-stability in \autoref{sec:stability}. 
We find that in the MSSM the issue of sufficient cosmological stability of the EW vacuum is subtle and quantitatively difficult to assess in the parts of parameter space which can account for the current muon $g-2$ discrepancy, although we present crude estimates. 
We also construct a non-minimal model, with additional simple vector-like Higgs superfields, in which it is possible to find examples where the EW vacuum is {\it absolutely} stable, consistent with the muon $g-2$ discrepancy, and therefore the thorny computation of vacuum lifetime is simply avoided.  The non-minimal model also includes regimes in which DM direct detection of bino DM is possible while {\it simultaneously} giving significant contribution to muon $g-2$, unlike the MSSM.

In \autoref{sec:benchmarks} and \autoref{sec:pheno}, we present and discuss various representative benchmarks in the MSSM as well as our non-minimal model, including collider phenomenology, their contributions to the muon $g-2$ and brief discussion of bino DM observability within direct detection experiments.
We also discuss the alternate interesting branch of $R$-parity violation. This becomes relevant if the DM arises from some more distant sector, such as axionic DM,
and we therefore  do not want a stable DM candidate within Sleptonic SUSY. This leads to a series of new searches that could lead to discoveries at the high-luminosity LHC.

In \autoref{sec:conclusions}, we conclude. In particular, we emphasize that multiple considerations converge to strongly motivate dedicated experimental searches for the superpartners of leptons and electroweak gauge bosons.

\section{Gaugino-Mediation} \label{sec:gauginomediation}

\subsection{Higher-dimensional Geography}

\begin{figure}[h]
\centering
\includegraphics[width=0.52 \linewidth]{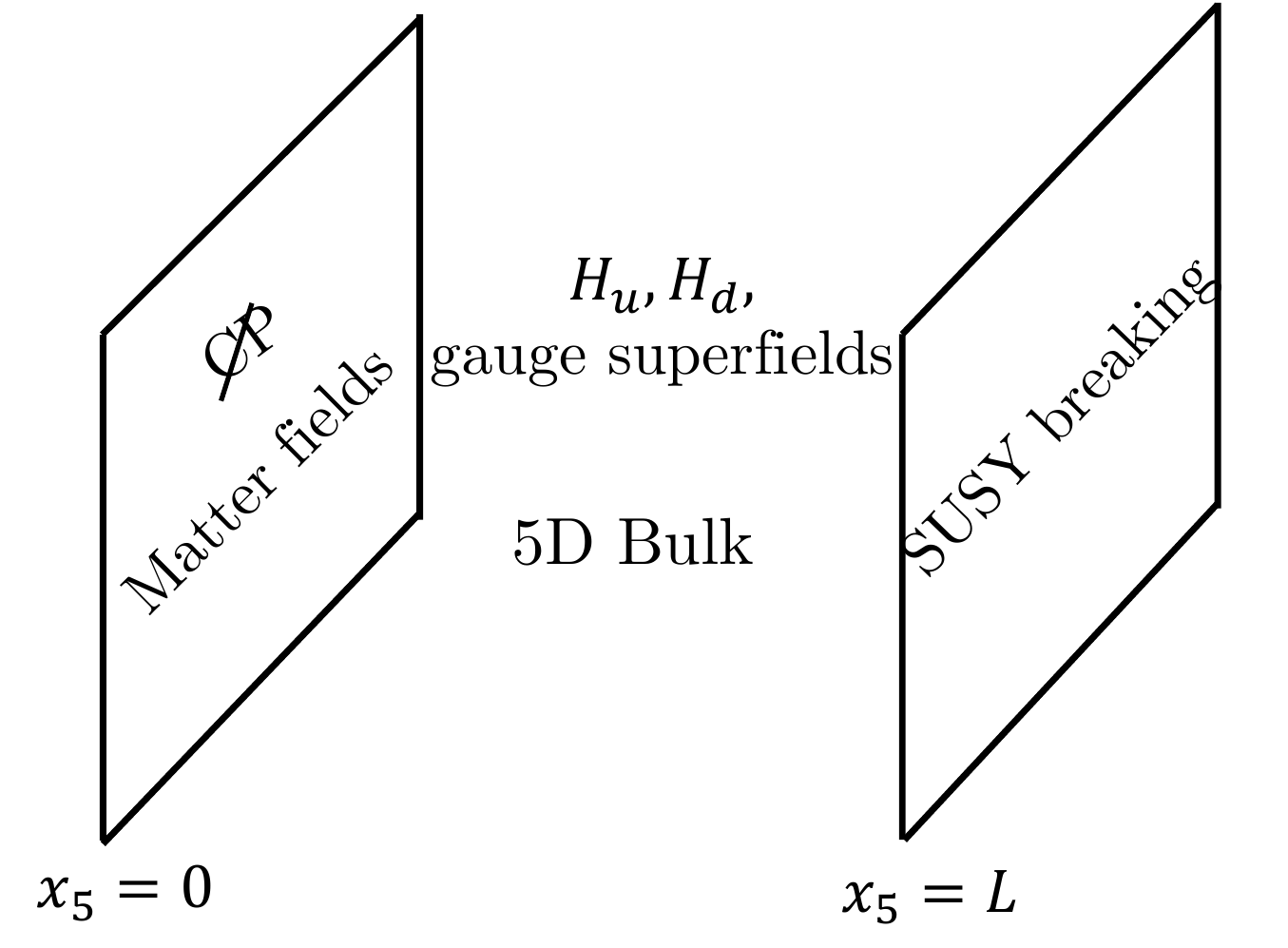}
\caption{UV structure of gaugino mediation. Flavored matter fields and Yukawa couplings and CP-violation are localized on the extra-dimensional boundary at $x_5 =0$, while SUSY-breaking dynamics is localized on the boundary at $x_5 = L$, thereby sequestering SUSY-breaking from flavor- and CP-violation.  }
\label{figure:gauginomed5D}
\end{figure}

The central UV structure is depicted in \autoref{figure:gauginomed5D}. Gaugino-mediation requires an extra-dimensional interval \cite{Chacko:1999mi, Kaplan:1999ac}, in which the SM fermions and sfermions (``matter'' superfields) are localized on one ($3+1$-dimensional) boundary, while the hidden sector responsible for spontaneous SUSY-breaking is localized on the other boundary. The gauge and Higgs superfields propagate in the $5$-dimensional bulk. Other bulk fields are taken to be significantly heavier than the Kaluza-Klein (KK) scale.
This extra-dimensional  ``sequestering'' of MSSM matter from SUSY-breaking means that at the KK scale at which we match to a $4$D EFT, there can only be negligible
flavor-violating SUSY-breaking for the sfermions. The dominant sources of flavor violation are therefore the Yukawa couplings of the Higgs superfields to the MSSM matter on the matter boundary. 
On the other hand, the bulk gauge and Higgs superfields can directly couple to the hidden sector, and this gives rise to SUSY-breaking effects for them at the KK scale. 

\subsection{SUSY-Breaking Structure}

The hidden sector is assumed to contain a singlet superfield $X$ which has a SUSY-breaking auxiliary field VEV (of mass dimension 2), $F_X \neq 0$. 
The relevant 4D effective Lagrangian (in terms of superfields) involving the MSSM then takes the form,
\begin{eqnarray} \label{eq:SUSYbreakingLag}
 \mathcal{L}_{ 4 \; \rm d } & \sim & \int d^2 \theta W_a^{ \alpha } W_{a \, \alpha } 
 \left( 1 + (c_{\lambda_a} \frac{X}{M} + \hbox{h.c.}) \right)  +  \\
& & \int d^4 \theta \Bigg\{  \left[ \left(1 + (c_{A_u} \frac{X}{LM^2} + {\rm h.c.})\right)H_u^{ \dagger } H_u + \left(1 + (c_{A_d} \frac{X}{LM^2} + {\rm h.c.})\right) H_d^{ \dagger } H_d\right] + \nonumber \\ 
&&\Big[ c_{\mu} \frac{X^{\dagger}}{ L \; M^2 } H_u H_d + \hbox{h.c.} \Big] 
+\frac{ X^{ \dagger } X }{ L \; M^3 } \Big[ c_{u} H_u^{ \dagger } H_u + c_d H_d^{ \dagger } H_d + \left( c_B H_u H_d + \hbox{h.c.} \right ) \Big] \Bigg\} \nonumber
\end{eqnarray}
where $W_{a, \alpha}$ is the gauge super ``field-strength'' 
associated with the MSSM gauge groups $a$,
$L$ is the length of the sequestering extra-dimensional interval and
$1/M$ typifies (non-renormalizable) couplings on the boundaries. In particular we take  the various dimensionless $c$ coefficients to be ${\cal O}(1)$,  varying by at most an order of magnitude from each other. 

For now, we will take $M L \sim {\cal O}(1)$ very roughly, and comment further on it and the compatibility  with successful sequestering 
in \autoref{sec:6D} when we take into account (orbifold) unification. 
Then we see that the terms depending on (the VEV of) $F_X$ clearly give soft masses at the KK scale, $m_{\lambda_a}, m^2_{H_u}, m^2_{H_d}, B \mu$, $A_u$, $A_d$
for the  4D effective MSSM $\sim c F_X/M$. 
We emphasize that
a distinctive virtue of $\rm \tilde{g}$MSB is that it both solves the SUSY flavor problem by not introducing any new flavor-violation in the soft masses, $m^2_{\rm sfermion} = 0$ at the KK scale, and simultaneously simply solves the SUSY $\mu$ problem via the Giudice-Masiero mechanism \cite{Giudice:1988yz},
realizing the supersymmetric $\mu$ 
parameter of the MSSM as a SUSY-breaking effect in the full theory, $\sim c_{\mu} F_X/M$. 
Note, the $c_{A_{u,d}}$ couplings give rise to $A$-terms at the KK matching scale (proportional to Yukawa matrices, satisfying the GIM mechanism). These were omitted in the original papers on ${\rm \tilde{g}mSB}$, but are not forbidden by any symmetries. They were studied in Refs. \cite{Brummer:2012ns, Heisig:2017lik}.
Finally, we do not include a bare $\mu$-term and a coupling of $X$ to $H_u H_d$ in the superpotential because these would  re-introduce a $\mu$ problem.
It is well known that such vanishing of terms in the superpotential are natural in SUSY, protected by non-renormalization theorems and symmetry(-breaking) superselection rules.

\subsection{Solution to the SUSY CP Problem}

The third major problem of the SUSY paradigm (indeed all BSM paradigms at the TeV scale), beyond the SUSY flavor and $\mu$ problems, is the SUSY CP Problem. Precision experiments very strongly constrain CP-violating electric dipole moments (EDMs), such as those of the electron or neutron, which however can easily get observable contributions from CP-violating phases in BSM physics at the TeV scale~\cite{Cesarotti:2018huy,Panico:2018hal}. Therefore  BSM theories that  naturally avoid having such new phases are more plausible, and this is a tight model-building constraint. In the SUSY context, this has to be accomplished consistent with a mechanism for solving the $\mu$ problem. Remarkably, in $\rm \tilde{g}$MSB this is straightforwardly achieved, by taking CP as a symmetry whose breaking is localized to the matter boundary of the higher-dimensional spacetime \cite{Chacko:1999mi}. This breaking then admits the observed CKM phase arising from the Yukawa couplings necessarily localized to the boundary with the matter superfields, while not introducing other CP-violation in soft SUSY breaking or the $\mu$ term originating from the other boundary.

The question of whether CP can be imposed as a fundamental symmetry, exact in some regions of the higher-dimensional spacetime, is a subtle and interesting one in a universe containing quantum gravity: see, for example, refs. \cite{Choi:1992xp}. But the experimental constraints on CP are so tight that it is worth looking at how $\rm \tilde{g}$MSB solves them in an alternate way, to assess its plausibility. One way to do this is to imagine that the higher-dimensional spacetime is moderately warped, so that there is a dual CFT/AdS interpretation (for a review, see, for example,  \cite{Aharony:1999ti}) along the lines understood \cite{Arkani-Hamed:2000ijo} for standard Randall-Sundrum I models \cite{Randall:1999ee}. If we take the matter boundary to be in the UV of the warped extra dimension and the SUSY-breaking to be localized on the IR boundary, then the dual interpretation is that matter supermultiplets are elementary fields coupled to a strongly-coupled 4D sector which dynamically spontaneously breaks SUSY in its IR \cite{Luty:2001jh}. The CP-breaking on the UV boundary with CP-conservation in the bulk and IR boundary then has the dual interpretation of ``accidental'' CP-conservation of the strongly-coupled sector, that is CP-violating effects are irrelevant in the strongly-coupled RG.

The question then becomes whether it is plausible that a strongly coupled 4D dynamics can naturally enjoy accidental CP symmetry in the IR. 
This would also seem to be a tough theoretical question, except that we have experimental/theoretical precedent from the strong interactions of SM QCD! We can think of (massless) QCD as a toy model, in which the renormalizable version is famously CP-symmetric. But if we consider its emergence from the far UV (say from string theory), we would also have non-renormalizable interactions in the effective IR theory, such as four-fermion operators, which in general would contain CP-violating phases. But these non-renormalizable interactions are of course rapidly negligible the further we flow into the IR, which is precisely the realization of accidental CP. We conclude that it is indeed plausible that a warped extra dimension realizes CP and localized CP-breaking in the manner described above, dual to this structure of accidental CP in 4D. By extension, this should be possible even when the warping is mild, as we take to be the case for the $\rm \tilde{g}$MSB framework.

\subsection{Gauge-coupling Unification and Gaugino Non-unification}

It is well known that the 4D MSSM RG gives a rather precise unification of gauge couplings at a  GUT scale $M_{\rm GUT} \sim 10^{15-16}$ GeV, strong evidence in favor of unification at that scale. In order to preserve this, we must take the sequestering KK scale $\geq M_{\rm GUT}$, so that 4D EFT operates all the way up to $M_{\rm GUT}$. Since our SUSY breaking masses are matched simply at the KK scale, we are necessarily in a 
high-scale SUSY-breaking scenario, where we must carefully track the large RG evolution of the above soft masses (and $\mu$-term) from KK to TeV scales.

Usually, the GUT structure is taken to give unified gaugino masses at $M_{\rm GUT}$, but the large MSSM running creates a modest hierarchy
in the IR, gluino $>$ wino $>$ bino \cite{Chacko:1999mi, Kaplan:1999ac}. 
But this is insufficient to realize the Sleptonic SUSY hierarchy. We therefore need a GUT structure compatible with non-universal gaugino masses already at $M_{\rm GUT} \sim M_{\rm KK}$, with variations within an order of magnitude.
 We will discuss such a GUT structure compatible with sequestering in \autoref{sec:6D}. But given such a structure, the RG flow can either counter the UV inequalities to yield comparable gaugino masses in the IR, or it 
 can accentuate the UV inequalities into sizeable hierarchies gluino $\gg$ wino $\gg$ bino. It is this latter robust option that leads in the IR to Sleptonic SUSY,  because in the RG, the heavy gluino drags up squarks to also be comparably heavy, while the lighter wino and bino are compatible with light sleptons. 

We will discuss the RG, first at one-loop order in the next section. A fuller account of orbifold GUTs, gaugino UV non-universality and sequestering is given in \autoref{sec:6D}.

\section{One-loop RGEs} \label{sec:OneLoopRGE}
As a first pass, in this section we restrict to the RGE to one-loop order 
\cite{Martin:1997ns}. We will use the notation of ref. \cite{Martin:1997ns} for the parameters and the RGE equations.

\subsection{Gauge-field and Gaugino Exchange}

Here, we will just focus on  the RG effects due to loops containing gauge-fields and gauginos.
One-loop RGE's for gauge couplings in MSSM [with $SU(5)$ normalization for hypercharge] are given by
\bea
\frac{ d \alpha_a^{ -1 } }{ d t } & = & - \frac{ b_a }{ 2 \pi }
\label{eq:beta_gauge}
\eea
where $t = \ln \mu$  with $\mu$ being the RG scale, $a = 1,2, 3$ denotes the 3 MSSM gauge groups and 
$\left( b_3, b_2, b_1 \right) =  \left( -3, +1, \frac{33}{5} \right)$.
We use $\alpha_1 ( 1 \; \hbox{TeV} ) \approx \alpha_1 \left( M_Z \right) \approx 0.017$ and $\alpha_2 
( 1 \; \hbox{TeV} ) \approx \alpha_2 \left( M_Z \right) \approx 0.034$
 and for QCD coupling, 
 first, we run $\alpha_3 \left( M_Z \right) \approx 0.118$ up to
$m_t \approx 173$ GeV using $b^{ \hbox{\small SM} - t }_3 = - 23/3$, resulting in $\alpha_3 \left( m_t \right) \approx 0.108$; then up to $\sim 1$ TeV using  $b^{ \hbox{SM} }_3 = - 7$
to give $\alpha_3 ( 1 \; \hbox{TeV} ) \approx 0.089$.
The RG equations for gaugino and sfermion masses are
\beq
\frac{d}{dt} M_a= \frac{1}{8 \pi^2} b_a g_a^2 M_a,
\eeq
\beq
16 \pi^2 \frac{d}{dt} m_{\phi_i}^2 =-\sum_a 8 C_a(i) g_a^2 |M_a|^2,
\eeq
where $C_a(i)$ are the quadratic Casimirs and $\phi_i$ is the scalar components of the superfield $\Phi_i$. The first of these equations governs the running of gaugino masses, from which $\frac{M_a}{\alpha_a}$ is found to be RG-invariant, giving
\beq \label{eq:gauginomass}
M_a({\rm IR})=\frac{\alpha_a({\rm IR})}{\alpha_a({M_{\rm GUT})}} M_a (M_{\rm GUT}),
\eeq
where the $\frac{\alpha_a({\rm IR})}{\alpha_a({M_{\rm GUT})}}$ ratios are $0.44$, $0.84$ and  $2.3$ for $a=1,2,3$ respectively. The second one is the RGE running of the other scalar soft masses sourced by the gaugino masses. 
Solving the RGEs, the scalar soft masses can be expressed in terms of the gaugino soft masses as in the following equations
\bea
m_{ {\phi_i} }^2 ( 1 \; \hbox{TeV} ) & = & 
m_{ {\phi_i} }^2 \left( M_{ \rm GUT } \right)  - 2 \sum_a 
C_a(i) \frac{ M^2_a ( 1 \; \hbox{TeV} ) }{ b_a } \Big[ 1 - \frac{ \alpha^2_a \left( M_{ \rm GUT } \right) }{ \alpha^2_a ( 1 \; \hbox{TeV} ) } \Big],
\label{eq:rge_sol_general}
\eea
and more explicitly for squarks and sleptons, for which $m_{ {\phi_i} }^2 \left( M_{ \rm GUT } \right)\approx 0$, we get
  \bea 
   m^2_{\tilde{Q}}\approx && m^2_{\tilde{u}}\approx m^2_{\tilde{d}}\approx (0.8 M_3 )^2, \nonumber \\ \label{eq:gauginomedsquarkmassall}
     m^2_{\tilde{L}} \approx && (0.44 M_1)^2 + (0.8 M_2)^2,   \\ \nonumber 
     m^2_{\tilde{e}} \approx && (0.88 M_1)^2,
    \eea
 where we have dropped the contributions $\propto M_{1,2}^2$ to squark masses, which can be neglected for $M_{1,2}\ll M_3$ in the IR.  
 So we see that the squarks obtain masses comparable to the gluino and the sleptons get masses comparable to bino and wino.

 Note that with just these effects, the singlet sleptons are lighter than the bino and so bino cannot be the LSP/dark matter \cite{Chakraborty:2019wav}.
   We will however discuss in the next subsection how the singlet sleptons can easily become heavier than bino, hence making the scenario of bino LSP/ co-annihilation dark matter possible in this model.
  
\subsection{$D$-mediation}  \label{subsec:Dmediation}

In addition to the  gauge-field and gaugino loops discussed above, there is another one-loop contribution controlled by gauge coupling. It comes from the Higgs loop (predominantly, as we will see later) and  from loops of other hypercharge carrying scalars with the hypercharge $D$-term coupling and is given by the following equation:
\beq \label{eq:dtermRGE}
16 \pi^2 \frac{dm^2_{\phi_j}}{dt} \supset \frac{6}{5} g_1^2 Y_j S,
\eeq
where $Y_i$ denotes the hypercharge and $S$ is defined as
\beq
S={\rm Tr}[Y_j m^2_{\phi_j}]=m^2_{H_u}-m^2_{H_d}+{\rm Tr} \left[{ \bf m^2_Q}-{\bf m^2_L}-2{\bf m^2_{\bar{u}}}+{\bf m^2_{\bar{d}}} +{\bf m^2_{\bar{e}}} \right].
\eeq
In ${\rm \tilde{g}}$MSB the soft masses for sfermions vanish in the UV, so $S$ is non-zero only if there is a mismatch between the Higgs soft masses in the UV. Moreover, it is easy to show that $\frac{S}{\alpha_1}$ is RG invariant at one loop. So if Higgs soft masses were equal in the UV, $S$ would stay zero under RG evolution and there would be no $D$-mediated contribution to scalar masses\footnote{This is however not the case at the two loop order. Starting with $S=0$ in the UV, two loop effects can generate nonzero $S$.}.

 Integrating eq.~\eqref{eq:dtermRGE}, we obtain the following expression for the hypercharge $D$-term contribution to the scalar soft masses
\beq
\Delta m_{\phi_i}^2 ({\rm IR})= \frac{Y_i}{ {\rm Tr}[Y^2]} \left( \frac{\alpha_1 ({\rm IR})}{\alpha_1({\rm UV})}-1 \right) S_{\rm UV}\approx 0.05 \, Y_i \, S_{\rm UV} ,
\eeq
where $S_{\rm UV}= {\rm Tr}\left[ Y_j m_{\phi_j}^2 \right]_{\rm UV}=m^2_{H_u}({\rm UV})-m^2_{H_d}({\rm UV})$.
This contribution to scalar soft mass squared is proportional to the hypercharge of the corresponding scalar. In particular note that the contribution has opposite signs for doublet vs. singlet sleptons due to their opposite sign of hypercharge. So if it becomes the dominant contribution, it would result in tachyonic masses for either the singlet or the doublet sleptons.
Therefore, in order for it to not be too large, we need $m_{H_u}^2 \approx m_{H_d}^2$, which is natural if the SUSY-breaking boundary approximately respects custodial symmetry.
We define 
\beq
\xi=\frac{m_{H_u}^2-m_{H_d}^2}{m_{H_d}^2}\Big|_{\rm UV},\label{eq:xidefinition4D}
\eeq 
which parametrizes the custodial symmetry breaking, and will take $\xi$ to be small.
%
If the slepton masses are hierarchically smaller than the soft Higgs masses (as in some of our benchmarks)
and in the absence of a hierarchy between squark and soft Higgs masses,
the $D$-term contribution to squark masses can be dropped, while the contribution to the slepton masses is important and is given by
\bea \label{eq:xiinslptonmassLR}
\Delta_\xi m_{\tilde{L}}^2 \approx (1.6)^2 \xi  \left(\frac{m^2_{H_d}}{10^2}\right), \\ 
\Delta_\xi m_{\tilde{e}}^2 \approx -(2.2)^2 \xi  \left(\frac{m^2_{H_d}}{10^2}\right).\nonumber
\eea
This contribution to the singlet slepton mass can easily raise the slepton masses above the bino \cite{Chacko:1999mi, Schmaltz:2000ei, Kaplan:2000av}, which as mentioned allows for having a bino LSP. Also in general it gives more flexibility in finding a variety of realistic spectra.

\subsection{One-loop Effects Controlled by Yukawa-couplings} \label{subsec:Yukawamediation}

In addition to the one loop RG effects controlled by the gauge couplings that we have discussed so far, Yukawa-coupling contributions can give important corrections to the above picture. First, the top Yukawa $y_{\rm t}$ plays a central role in driving radiative EWSB through the following term:
\beq
16 \pi^2 \frac{dm^2_{H_u}}{dt} \supset 6 |y_t^2| (m^2_{H_u}+m^2_{Q_3}+m^2_{\bar{u}_3})+6|a_t|^2.
\eeq
Similarly, we have the RGE for $m^2_{H_d}$:
\beq
16 \pi^2 \frac{dm^2_{H_d}}{dt} \supset 6 |y_b^2| (m^2_{H_d}+m^2_{Q_3}+m^2_{\bar{d}_3})+6 |a_b|^2.
\eeq
where we take into account one-loop threshold corrections to the bottom quark mass in computing the IR value of $y_b$ \cite{Carena:1999py}. The $A$ terms run according to the following equation.
\bea \label{eq:atermRGE}
16 \pi^2 \frac{d a_t}{dt}= && a_t \left[ 18|y_t|^2+|y_b|^2-\frac{16}{3}g_3^2-3g_2^2 -\frac{13}{15}g_1^2 \right]+2 a_b y_b^* y_t  \\ \nonumber 
&& +y_t \left[\frac{32}{3}g_3^2 M_3 + 6 g_2^2 M_2 +\frac{26}{15}g_1^2 M_1 \right] \\ \nonumber
16 \pi^2 \frac{d a_b}{dt}= && a_b \left[ |y_t|^2+18|y_b|^2+|y_\tau|^2-\frac{16}{3}g_3^2-3g_2^2 -\frac{7}{15}g_1^2 \right]+2 a_t y_t^* y_b + 2 a_\tau y_\tau^* y_b \\ \nonumber 
&& +y_t\left[\frac{32}{3}g_3^2 M_3 + 6 g_2^2 M_2 +\frac{14}{15}g_1^2 M_1 \right] \\ \nonumber 
16 \pi^2 \frac{d a_\tau}{dt}= && a_\tau \left[ 3|y_b|^2+12 |y_\tau|^2-3g_2^2 -\frac{9}{5}g_1^2 \right]+6 a_b y_b^* y_\tau   +y_\tau \left[ 6 g_2^2 M_2 +\frac{18}{5}g_1^2 M_1 \right]
\eea

Moreover, $y_{\tau}$ with large $\tan \beta$ and large Higgs soft terms can significantly affect the stau spectrum. 
The following equations show the one loop contribution, with Higgs in the loop, to the running of doublet and singlet stau masses controlled by $y_\tau$,
\bea \label{eq:RGEHiggsforL3E3}
16 \pi^2 \frac{dm^2_{L_3}}{dt} \supset 2 |y_\tau^2| m^2_{H_d}+2|a_\tau|^2 \\  
16 \pi^2 \frac{dm^2_{\bar{e}_3}}{dt} \supset 4 |y_\tau^2| m^2_{H_d}+4|a_\tau|^2.\nonumber
\eea
The related $A$-terms run according to eq.~\eqref{eq:atermRGE}.
A positive (negative) $m^2_{H_d}$ gives negative (positive) contribution to slepton mass squared. Given the hierarchy between slepton and Higgs masses, a large positive $m^2_{H_d}$ could result in too light or even tachyonic staus. 
On the other hand, as we will discuss in more detail in later sections, in order to achieve an attractive and viable Sleptonic SUSY spectrum, the choice of negative $m^2_{H_d}$ to raise the stau mass is essential. In what follows, we will refer to this effect as $y_\tau$-mediation or more simply as Yukawa-mediation, and to all the contributions to sfermion masses $\propto m^2_{H_{u,d}}$ generally as Higgs-mediation.

In addition to the one-loop RGEs considered so far, there are two-loop effects that are important quantitatively, but not qualitatively in the regime we will consider. We will therefore postpone these effects until \autoref{sec:twoloopRGE}.

\section{Muon Magnetic Moment} \label{sec:bottomup}

SUSY, and in particular Sleptonic SUSY, provides an attractive framework for addressing the potential muon $g-2$ anomaly,
for a number of sharp reasons. 
To appreciate this, we begin {\it without positing SUSY}, just assuming that the new physics responsible for the $g-2$ discrepancy is weakly-coupled, therefore predominantly appearing within one-loop diagrams. 
Such loops must then consist of a fermionic line and a bosonic line, the latter being either spin-$1$ or spin-$0$ in perturbative field theory. In general, a massive spin-1 $Z'$ gauge boson is an attractive BSM possibility if its fermionic couplings are universal across all generations, because it enjoys an elegant extension of the SM GIM mechanism for suppressing 
%
%
FCNC's. However, it is extremely difficult for such a $Z'$ to account for the muon $g-2$ discrepancy without being excluded by searches that relate the couplings to muons and electrons \cite{Athron:2021iuf}.  Non-universal $Z'$'s can evade such search constraints, but are also severely constrained by various searches~\cite{Athron:2021iuf,Amaral:2021rzw}, and avoid lepton FCNCs, such as $\mu \rightarrow e + X$, by assigning the three generations different $Z'$ charges so that their Yukawa matrices and $Z'$ interactions are necessarily flavor-diagonal in the same basis (with neutrino masses/mixing arising as a small correction to this). But in the SM, the GIM mechanism remarkably protects against FCNCs even when there is no distinction between generations in the gauge assignments and the Yukawa matrices are a priori non-diagonal. A $Z'$ with generation-dependent charges would then make the elegant GIM structure of the SM a pure accident, rather than a deep theme of physics. Here, we consider maintaining the GIM mechanism as a guiding principle for any BSM extension underlying the $g-2$ discrepancy, in particular allowing the three generations to have identical (BSM) gauge quantum numbers, and hence general Yukawa matrices, as for quarks.
Generation-dependent $Z'$ extensions are therefore less attractive from this viewpoint.

On the other hand, new spin-$0$ bosonic lines bring their own challenges, both in evading excessive FCNCs as well as their {\it own} hierarchy problem, beyond that of the SM Higgs boson. Famously, SUSY solves the Higgs hierarchy problem as well as the hierarchy problem of scalar superpartners, so the latter are good candidates to contribute to $g-2$. But unlike gauge bosons, where universal minimal couplings among generations is field-theoretically robust, there is no such universal coupling structure for scalars. So in general, spin-$0$ couplings to leptons will be flavor-dependent. In order to maintain the GIM mechanism, this BSM flavor-dependence must somehow be given by copies of the Yukawa matrices, which is difficult to realize robustly.\footnote{Minimal Flavor Violation (MFV) \cite{DAmbrosio:2002vsn} is often invoked in general BSM settings as a structure of this type, with the Yukawa matrices replicated in the BSM couplings. But  MFV is itself just an ansatz, that must then be supplemented by an explanatory dynamical mechanism.} In the SUSY paradigm however, where the scalars are sleptons, the copying of the Yukawa matrices is economically and automatically enforced by supersymmetry itself. SUSY-breaking soft masses {\it can} however introduce new sources of flavor violation and spoil the GIM mechanism, but remarkably again there are 
 a number of elegant SUSY-breaking field theory mechanisms, such as
 GMSB, $\tilde{g}$MSB
 and AMSB 
 which suppress new flavor-violation. For the reasons discussed earlier we are specifically focused on $\rm \tilde{g}$MSB.
 But the general feature is that SUSY allows us to only have the Yukawa matrices as sources of flavor-violation, and therefore the GIM mechanism can continue to hold in SUSY extensions of the SM. 
Therefore, it is broadly plausible that sleptons mediate sizeable $g-2$ contributions. 

We can estimate that loop contributions to the effective $g-2$ of any of the leptonic flavors, integrating out new particles with very roughly order-one (or smaller) couplings, will have the general form,
\beq
{c_\ell} \bar \ell_L \sigma_{\mu\nu} 
\ell_R F^{\mu\nu} H,~c_\ell\sim \frac {\kappa y_\ell} {M^2_{\rm BSM}} 
\eeq
where $\Delta a_{\ell}\equiv c_\ell~\! m_\ell~\!v$ is the conventionally normalized BSM shift in the leptonic magnetic moment. To account for the current muonic discrepancy, we would require $\Delta a_{\mu} = 
259 \pm 60 \times 10^{-11}$. Here, 
$v \sim 174$ GeV, $\kappa$ is a product of flavor-blind, 
dimensionless
couplings, and $y_{\ell = e, \mu, \tau}$ is the diagonalized Yukawa matrix, the presumed sole source of chirality breaking and flavor-violation. To have not been observed yet, we must take $M_{BSM} \gsim v$. To be weakly-coupled, the strongest such effect, for $\tau$, should satisfy
$\kappa y_{\tau} \lesssim {\cal O}(1)/(16 \pi^2)$.
Using the scaling $c_\ell \propto y_\ell$,
this implies
\beq
\Delta a_\mu \lesssim \mathcal{O}(1)\times 10^{-8} \left(\frac {\TeV} {M_{\rm BSM}}\right)^2.
\label{eq:generalbound}
\eeq

We see that to account for the existing anomaly, we need to approach this conservative bound. Note that this is quite non-trivial, because it appears this requires the $\tau$ $(g-2)$ to have no parametric suppression (beyond the $\sim 1/(16 \pi)^2$), whereas naively one might have expected at least an 
additional
suppression by the SM $\tau$ Yukawa coupling $\sim 10^{-2}$. 

  \begin{figure}[h]
\centering
\includegraphics[width=0.5 \linewidth]{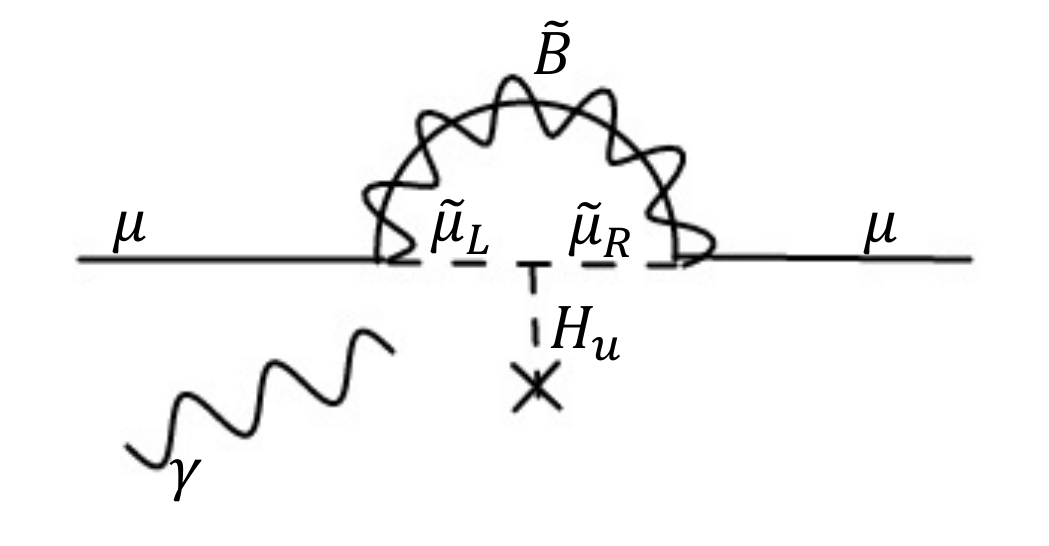}
\caption{A SUSY Feynman diagram contributing to the muon $g-2$, enhanced for large $\mu$ and $\tan \beta$. The photon line may be attached to either of the smuons.}
\label{figure:FeynmandiagBinosmuon}
\end{figure}

But SUSY can compensate such an expected suppression 
in 
$\tau$ $g - 2$ 
(hence $\Delta a_{ \mu}$),
as can be seen in one of its contributions, in \autoref{figure:FeynmandiagBinosmuon}. There, the chirality-breaking needed for $g-2$ appears within a supersymmetric ($F$-term potential) trilinear scalar coupling of the left- and right-handed sleptons to $H_u$. In the SM, $y_{\ell}$ is inferred from the observed masses and EW scale $v$, but SUSY  (minimally) has two Higgs doublets $H_{u,d}$  so that the $y_{\ell}$ must be inferred from the masses and $v_d$. This gives stronger Yukawa couplings for large $\tan \beta \equiv v_u/v_d$, with $v_u \approx v$, which readily occurs for moderately small Peccei-Quinn breaking soft parameter $B \mu$, 
\beq \label{eq:tanbeta}
\tan \beta\approx\frac{m^2_{H_d}+ |\mu|^2-m_Z^2/2}{B \mu}.
\eeq
In terms of SM Yukawa couplings, the trilinear couplings are then  $y^{\rm SM}_i \mu\; \tan \beta$. The {\it dimensionless} strength of such a trilinear coupling is given by $y_i \mu \tan \beta/m_{\tilde{\ell}_i}$.  We therefore see that the Yukawa-coupling suppression can be compensated by taking moderately large $\tan \beta \sim {\cal O}(10)$ as well as large $\mu \sim {\cal O}(10) m_{\tilde{\ell}_i}$,
which combine to 
roughly
overcome
the anticipated suppression of $y_{ \tau }$ mentioned above.
In such a regime, we remain weakly-coupled while approaching the bound of eq.~\eqref{eq:generalbound}. It is attractive that the requisite two-Higgs doublet structure,
allowing $\tan \beta$ enhancement,  is automatic within SUSY extensions of the SM.

Large $\tan \beta \sim {\cal O}(10)$ is standard in SUSY parameter space, maximizing the Higgs boson mass at tree-level, which is a first step in fitting the 
 ``large'' (by SUSY standards) observed Higgs boson mass of $125$ GeV. 
 This
endows $H_u$ with approximately SM-like properties, while allowing the $H_d$ Yukawa couplings to remain perturbative. But for weak-scale sleptons, $\mu \sim {\cal O}(10) m_{\tilde{\ell}}$ is in the multi-TeV range, which is less familiar. However, this also fits quite plausibly with the large loop-level contributions
needed for the $125$ GeV Higgs boson mass (that is, ``large'' contributions to the quartic coupling of $H_u$),  and most straightforwardly accounted for by having EWSB triggered radiatively by multi-TeV stop masses \cite{Draper:2011aa}. A prototypical model of this kind is ``Mini-Split SUSY'' \cite{Arvanitaki:2012ps, Arkani-Hamed:2012fhg}, in which {\it all} BSM scalars $\sim {\cal O}(10)$ TeV, while gauginos are lighter. Here, we will require all the squarks to be heavy $\sim {\cal O}(10)$ TeV, but for the sleptons to be light $<$ TeV so that they contribute significantly to the muon $g-2$. This remains consistent with an extended GIM mechanism protecting against excessive FCNCs, but poses an interesting challenge for SUSY breaking dynamics to realize this novel hierarchy while maintaining GIM. 
The plausibility of multi-TeV $\mu$ can be assessed by noting that the constraint for successful EWSB in the MSSM (at $\tan \beta \sim {\cal O}(10)$, for parameters at EW-scale) reads
\beq
-m_Z^2/2  \approx \mu^2+m_{H_u}^2 + {\rm loop ~and ~1/\tan \beta ~corrections.}
\label{eq:EWSB}
\eeq
The heavy stop masses $\sim {\cal O}(10)$ TeV
 imply that $|m^2_{H_u}|$ is {\it naturally}  
$\sim {\cal O}(10 {\rm TeV})^2$ 
 as well, given the large top Yukawa coupling (and the large RG logarithms in high scale SUSY breaking models such as $\rm \tilde{g}$MSB). Therefore, the smallness of the RHS of eq.~\eqref{eq:EWSB} can be achieved by fine-tuning $m^2_{H_u}({\rm IR}) \sim m_Z^2$ and also taking $\mu \sim m_Z$, {\it or} simply taking $\mu \sim 10$ TeV but and tuning it against 
$m^2_{H_u}({\rm IR}) \sim {\cal O}(10)$ TeV to give the small RHS. Once we accept this modest fine-tuning necessary in either option (which is the SUSY face of the little hierarchy problem), we see that the second choice of $\mu \sim {\cal O}(10)$ TeV is robust. 
All these considerations show us why taking the muon $g-2$ discrepancy seriously leads to Sleptonic SUSY as a prime explanatory framework.

Committing to a SUSY explanation of the muon $g-2$ discrepancy in terms of slepton/bino exchange, we have~\cite{Moroi:1995yh,Endo:2021zal}
\bea
\Delta a_\mu &=& \frac {a_Y} {4 \pi} \frac {\mu M_1 m_\mu^2} {m_{\tilde \mu_L}^2 m_{\tilde \mu_R}^2} \tan\beta \times \frac 1 {1+\Delta_\mu}\times f_N\left(\frac {m_{\tilde \mu_L}^2}{M_1^2}, \frac {m_{\tilde \mu_R}^2}{M_1^2}\right) \label{eq:g2}\\
&\simeq& 200\times 10^{-11} \left(\frac {\mu} {10~\TeV}\right) \left(\frac {(500~\gev)^3 M_1}{m_{\tilde \mu_L}^2 m_{\tilde \mu_R}^2}\right) \left(\frac {\tan\beta} {20}\right) \left(6 f_N\left(\frac {m_{\tilde \mu_L}^2}{M_1^2}, \frac {m_{\tilde \mu_R}^2}{M_1^2}\right)\right). \nonumber
\eea
Here $\Delta_\mu$ is the radiative correction to the muon mass (which is fixed in the IR). This correction can be as large as $\mathcal{O}(10\%)$ for a typical benchmark considered in our study, which we have taken as a default in the second line.
Note that the loop function $f_N\left(\frac {m_{\tilde \mu_L}^2}{M_1^2}, \frac {m_{\tilde \mu_R}^2}{M_1^2}\right)$~\cite{Endo:2021zal}
\beq
f_N(x,y)=xy\left[\frac{-3+x+y+xy}{(x-1)^2(y-1)^2}+\frac{2x\ln x}{(x-y)(x-1)^3}-\frac{2y\ln y}{(x-y)(y-1)^3}\right],
\eeq
has a ``typical'' value of $1/6$ for $f_N(1,1)$ but in general it scales with these mass ratios differently in different regimes.

The detailed sleptonic spectrum is central to the collider phenomenology as well as the $g-2$ effects, so let us consider its structure. As discussed above, in order to maintain the GIM mechanism we need the patterns of SUSY breaking to only depend on flavor-blind effects such as those due to gauge couplings, with any flavor-violation proportional to the Yukawa coupling matrices. Since the Yukawa couplings are necessarily very small for the selectrons and smuons, we expect the mass eigenstates to be approximately the gauge eigenstates $\tilde{\ell}_L$ and $\tilde{\ell}_R$, and each approximately degenerate between the first two generations. 
While these $L/R$ eigenstates may be split from each other, they should be $\lsim 500$ GeV to account for the muon $g-2$. Sneutrinos will be approximately degenerate with $\tilde{\ell}_L$. The $\tau$ Yukawa is however large enough to significantly impact the stau spectrum, as is already clear from the trilinear coupling to the Higgs discussed above, ${m_{\tau}\over v} \mu \tan\!\beta~\tilde{\ell}_{3 L} H_u \tilde{\tau}$. 
Since we have chosen this dimensionful coupling to be $\sim {\cal O}(m_{\tilde{\ell}}) \sim {\cal O}(m_{\tilde{\ell}})$, so as to maximize the perturbative $g-2$ SUSY corrections, we see that upon EWSB it yields a $L/R$ mixing comparable to the slepton masses themselves. The stau mass eigenstates $\tilde{\tau}_{1,2}$ will therefore be significant admixtures of the gauge eigenstates, and have significantly different mass eigenvalues from the selectrons and smuons. It is important of course that the lightest eigenstate $\tilde{\tau}_{1}$ is not too light for existing constraints, or even tachyonic.

\section{One-loop Considerations for Sleptonic SUSY} \label{sec:1loopconsiderations}

In this section, we will piece together the gaugino-, D- and Higgs-mediated SUSY breaking  that preserves the GIM mechanism, arising from the sequestered UV structure, 
and show how the RGE at one-loop order can fit these with the IR phenomenological considerations. While, it will be ultimately necessary to  go to two-loop order to get a reasonable approximation to the IR spectrum, the one-loop description gives a simple yet qualitatively accurate account of the main considerations. With one-loop RGE, we can use tree-level matching of the 5D theory to the 4D EFT at the KK matching scale:
$m^2_{\rm sfermion}({\rm UV}) \approx 0$, $\alpha_a ({\rm UV}) = \alpha_{\rm unified}$ and , $m^2_{H_u}({\rm UV}) \approx m^2_{H_d}({\rm UV})$.
While UV $A$-terms are allowed, they are most important in the stop/Higgs sector, where if they are large enough they can lower the stop mass compatible with a $125$ GeV Higgs. However, we find in the current framework this does not significantly improve the electroweak fine-tuning of the LHP. Also, there is  limited parameter space remaining for light stops (or gluinos) to be discovered at the LHC, given the existing constraints. Therefore,  we will 
study the regime of modest $A_{u,d}({\rm UV})$, and neglect their effects in the remainder of the paper.

We begin by considering the slepton and squark masses due to just gaugino-mediation, eqs. \eqref{eq:gauginomedsquarkmassall}. Clearly, to have heavy stops (and other squarks) $\sim 10$ TeV, with $\sim$ weak-scale sleptons, requires hierarchical gaugino masses, $M_3 \sim {\cal O}(10)$ TeV, $M_{1,2} < $ TeV, in the IR. 
 The multi-TeV $\mu$ discussed above will dominate the Higgsino masses, and therefore they are phenomenologically irrelevant at the LHC, just like the squarks, gluino, and the $H_d$ scalars. Furthermore, large $\mu$ implies that there is very little mixing with the bino and wino after EWSB, so that these gauginos are essentially the mass eigenstates. Note, the very heavy gluino easily avoids the very stringent LHC search constraints 
$\gsim 2$ TeV. However, the winos are non-trivially constrained by LHC searches to be $\gsim$ TeV (for uncompressed sleptonic spectra). Phenomenologically, it is possible that they may still be discovered at the LHC if they are just above current bounds, but they may readily be out of reach as well.
In this paper, we will first consider the simplest (and flavor/CP safest) R-parity conserving option for SUSY, in which case the bino can serve as the lightest supsersymmetric particle (LSP). Being stable by R-parity, it then lies at the bottom of all supersymmetric decay chains and escapes the detectors invisibly. Famously, it can also be the primary constituent of dark matter, with a thermal relic abundance matching observations if it dominantly co-annihilates with the lightest slepton in the early universe. This requires the bino to be only a few 
GeV lighter than this slepton.

We see that we need a gaugino hierarchy in the IR with weak-scale bino LSP, $\sim$ TeV-scale wino and $\sim$ 10 TeV gluino.
While some of this IR hierarchy arises in running from the UV, eq.~\eqref{eq:gauginomass}, clearly we must have started with a hierarchy, $c_{\lambda_3}/c_{\lambda_1} \sim {\cal O}(10)$ already in the UV, eq.~\eqref{eq:SUSYbreakingLag}. 

In more detail, we see that pure gaugino-mediation would be incompatible with our desired spectrum, because the right-handed sleptons are predicted to be lighter than the bino in that case, and the left-handed sleptons would have comparable mass to the $\sim$ TeV wino which would strongly suppress muon $g-2$ contributions. Fortunately, both of these issues can be resolved by invoking the $D-$mediated effects, eqs.~\eqref{eq:xiinslptonmassLR}~\cite{Chacko:1999mi, Schmaltz:2000ei, Kaplan:2000av}, for $m_{H_{u,d}}^2 \sim (10 \; {\rm TeV})^2$ and for $\xi \sim$ few percent.  Fundamentally, by tree-level matching, $\xi = (c_u - c_d)/c_d$ from eq. \eqref{eq:SUSYbreakingLag} and eq.~\eqref{eq:xidefinition4D}. Given that $c_u = c_d$ would be a consequence of custodial $SU(2)$ symmetry on the SUSY-breaking boundary, we are assuming that this is an approximate symmetry there, which is natural. 

The final important qualitative consideration relates to the stau spectrum. The MSSM contains a supersymmetric scalar trilinear coupling ${m_{\ell}\over v} \mu \tan\!\beta~\tilde{\ell}_L H_u \tilde{e}$ arising from the Yukawa couplings and $\mu$ term in the superpotential.  
When the sleptons here are smuons, this coupling contributes significantly to the muon $g-2$, eq.~\eqref{eq:g2}, but is otherwise a modest coupling for collider purposes, given the small size of $m_{\mu}/v$. However, for the large  $\mu \sim {\cal O}(10)$ TeV and $\tan \beta \sim {\cal O}(10)$ we have been led to consider, this trilinear coupling for staus, 
${m_{\tau}\over v} \mu \tan\!\beta~\tilde{\ell}_{3 L} H_u \tilde{\tau}$ is significant. For example, after EWSB, this yields stau mixing, $\sim {\cal O}(1/2~{\rm TeV})^2$.
If the EW-invariant slepton mass$^2$s are universal and $\sim v^2$, then after  diagonalization the lighter stau eigenstate will be tachyonic. To avoid this, we need non-universal (but flavor-preserving) EW-invariant mass$^2$ such that the staus are somewhat heavier, which can arise through the Yukawa-coupling effects. The relevant RG effects are reviewed in \autoref{subsec:Yukawamediation}.  As pointed out there, despite the small leptonic Yukawa couplings such effects become important when the Higgs soft masses are large, precisely as occurs in Sleptonic SUSY. In particular, the staus can be made heavier if $m_{H_d}^2 < 0$ with large magnitude.\footnote{Note, this does not imply a large $H_d$ condensate as long as 
$|\mu|^2 + m_{H_d}^2 > 0$, as we take to be the case.}
We see that for $\sim (10 \; {\rm TeV})^2$ Higgs soft terms and $\tan \beta \sim 10$, the solutions to the RGE, eqs.~\eqref{eq:RGEHiggsforL3E3}, can give stau-smuon(selectron) EW-invariant mass splittings sufficient to avoid stau tachyons.

There is a related but subtle requirement that we will discuss further in 
 \autoref{sec:stability}. Even avoiding tachyonic staus as above,
  the sizeable stau-Higgs trilinear coupling will typically give rise to a deeper minimum of the scalar effective potential, in which the staus will still condense along with the Higgs, thereby breaking electromagnetism! One then has to ensure that the preferred local minimum in which only the Higgs condenses is at least metastable. The tunneling decay rate of this  preferred metastable vacuum is non-perturbatively weak in the couplings
  , but it still places a stringent requirement that the resulting lifetime is cosmologically long. This requires even larger EW-invariant stau masses than for simply avoiding tachyons after mass-diagonalizaton, but can still be achieved via Yukawa-coupling effects. 
  
  While the various considerations above are easiest to understand at the one-loop level, there are significant two-loop effects due to the hierarchical spectrum that should be included in constructing accurate benchmarks. But these will not change the qualitative features. We will discuss these in sections \autoref{sec:twoloopRGE} and \autoref{sec:benchmarks}.

\section{``Living Dangerously'' and Frustrated Naturalness} \label{sec:livingdanerously}

There appear to be two uncomfortable aspects to Sleptonic SUSY. The first, as 
 discussed above is that Sleptonic SUSY is dangerously close to a transition to a vacuum in which electromagnetism is Higgsed by stau condensation.  The second is that 
given the tie between SUSY and naturalness, it raises the burning question of why nature would have chosen a structure such as Sleptonic SUSY which suffers from 
the 
%
%
LHP, 
rather than 
 squarks having shown up naturally already at the weak scale? In this section, we present a scenario in which these two aspects may be correlated with each other.
 
 The whole issue of naturalness in general has been reconsidered in light of the proclivity of complex field theories and string theories to possess a multitude of (meta-)stable vacuum solutions. Combined with General Relativity, these can appear as a large multiverse in which sub-universes are governed by different EFTs, and these can appear multiple times with different effective parameters. In such a framework, there can be a tension between naturalness of an EFT realization and anthropic criteria. That is, in some classes of EFTs full field theoretic naturalness may not hold in the part of parameter space in which the appearance of intelligent lifeforms (such as ourselves) is viable. If one starts in such an anthropically viable region of parameter space, one can allow the EFT to become more and more natural until arriving at the boundary at which the EFT is at the cusp of violating an anthropic criterion. One would then expect to be living in one of these most natural (but not fully natural) EFTs at this anthropic boundary in the space of EFTs.  
  This expectation has been called the 
 ``principle of living dangerously'' \cite{Weinberg:1987dv, Arkani-Hamed:2005zuc}. In the SUSY EW context it was first invoked in  ref. \cite{Giudice:2006sn}. 
 Of course a difficulty in checking this principle in a particular EFT is to identify what is a clear anthropic criterion. A conservative approach is to ask if the EFT is close to a ``phase transition'' in parameter space, since this is more unambiguous, and very plausibly crossing into the ``wrong'' phase could dramatically lower anthropic viability.
 
 The classic example of living dangerously in this sense is Weinberg's analysis of the cosmological constant problem \cite{Weinberg:1987dv}, in which he showed that the standard GR, 
 plus
 %
%
matter EFT with realistic parameters has a cosmological constant which is near the boundary  
 between the regime (``phase'') in which galaxies can form. If the cosmological constant were larger and more natural, galaxies would not form. In this analysis, there is a massive contrast between pure naturalness and naturalness tempered by the plausible anthropic criterion of galaxy formation. But in other situations, there may be a milder anthropic compromise of pure naturalness considerations, such as in the Little Hierarchy Problem, which we will refer to as Frustrated Naturalness
 \cite{Ramantalk}.

 There is an apparent ambiguity in Weinberg's analysis because he considers varying the cosmological constant 
  in testing for galaxy formation, {\it while keeping other parameters fixed} (such as the initial size of density perturbations). But one can generalize this analysis and allow for {\it small changes} in all input parameters, because only the IR effective cosmological constant is very sensitive to these. We will refer to this as a ``local'' test of living dangerously.
   Beyond this, any type of major ``global'' move in parameter space is notoriously difficult to assess in terms of realizability within the multiverse and in terms of anthropic viability. Of course, the local test is not definitive in proving that an apparent breakdown of naturalness is due to anthropic ``pressure'' in a multiverse, but it is at least a modest piece of circumstantial evidence in that direction.
 
 We will see that, remarkably, Sleptonic SUSY is ``living dangerously'' in this local sense. 
 First consider a realistic Sleptonic SUSY spectrum along the lines discussed  and motivated in the sections above, with sleptons $\sim$ EWSB scale but with $\mu$, squark and Higgs soft masses $\sim {\cal O}(100)$ times heavier, and with $\tan \beta \gsim 10$. 
 Of course, the large soft terms  imply a LHP with  $\sim 10^{-4}$ EW fine-tuning, eq.~\eqref{eq:EWSB}.  But we can ask what would happen if we de-tune from this point in MSSM parameter space by allowing all parameters to vary at, say, the $\sim 10^{-3}$ level, so as to make the EW fine-tuning milder $\sim 10^{-3}$. Generally, this would perturb all features of the physical spectrum at this tiny level, 
 {\it except} for the finely-tuned EW scale.  The two IR soft terms on the left-hand side eq.~\eqref{eq:EWSB} are ${\cal O}(10 \,{\rm TeV})^2$ before and after de-tuning, but before de-tuning they finely cancel to within $\sim 
{\cal O}(100 \,{\rm GeV})^2$, while after de-tuning they will cancel far less.  
 One distinct possibility is that the left-hand side of eq.~\eqref{eq:EWSB} may now flip sign, so that EWSB no longer takes place. This clearly is a phase transition.\footnote{In this case, EWSB still takes place due to QCD chiral symmetry breaking, but in such a dramatically different regime that we can still think of it as a phase transition in which anthropic viability is likely different, and plausibly much worse.} But this much is true of any broadly natural extension of the SM which suffers from the
 LHP.
 %
 %
The other possibility is that after de-tuning, the left-hand side of eq.~\eqref{eq:EWSB} remains negative, resulting in substantially larger $m_Z \sim 400$ GeV or the weak scale $v \sim 800$ GeV. This is less fine-tuned ($\sim 10^{-3}$) but in Sleptonic SUSY it triggers a transition into a far less anthropically viable phase of a massive photon! 

To see this, note that  unlike the EW-symmetric soft masses, the stau-mixing, $y_{\tau} \mu v \; \tilde{\tau}_L  \tilde{\tau}_R$, is sensitive to the altered scale of EWSB, $v$. We are considering  $\tan \beta \sim 10$, so that 
$y_{\tau} \sim 10 \; y_{\tau}^{\rm SM} \sim 0.1$, and $\mu \sim {\cal O}(10)$ TeV, so the stau mixing mass$^2 \sim v$.TeV. Given EW-symmetric stau masses from $y_\tau$-mediation $\lesssim$ TeV, eqs.~\eqref{eq:RGEHiggsforL3E3}, which are insensitive to the de-tuning, the substantially larger $v$ will result in the lighter mass eigenstate stau becoming tachyonic, so that it condenses and Higgses EM.
This is the sense in which Sleptonic SUSY is ``living dangerously'' \cite{Weinberg:1987dv, Arkani-Hamed:2005zuc}: modest variation of parameters  that 
substantially improves the naturalness of the model inevitably leads to a phase transition.
In a large multiverse of MSSMs which sample this local vicinity of parameter space, the only ones in which the standard (presumably more anthropically viable) phase arises are 
 those in which the theory ``suffers'' from a significant LHP. Again, we can only reliably perform this local analysis, we cannot make any conclusion about large deviations in the parameter space and how prevalent these are in the multiverse. 
 
 For even the  rather limited version of ``living dangerously'' to apply, it is obviously important that there are no small ``local'' variations in parameter space which dramatically improve naturalness and yet evade the phase transitions to no EWSB or to the massive photon. There are 
 two subtleties in this regard, that the large stau-mixing needed to trigger the massive-photon phase might be significantly reduced in the local variations, or that when Higgs condensation is shut off there might be another source of significant EWSB that turns on due to local variations. Let us consider each case. The large size of the stau-mixing mass-squared is due to two enhancements: large $\mu \sim {\cal O}(10 {\rm TeV})$ and large $\tan \beta \sim {\cal O}(10)$. Since we are allowing local variations of the fundamental parameters of our theory at the $\sim 10^{-3}$ level we should check that we cannot dramatically reduce these enhancements by such local variations, and thereby avoid the massive-photon transition by stau-condensation. Now, $\mu$ is itself one of these fundamental parameters, so local variations will not change it appreciably, but  $\tan \beta$ is a derived quantity and we should therefore check. For large $\tan \beta$ and large $m_{H_d}^2$, we have $\tan \beta\approx\frac{m^2_{H_d}+ |\mu|^2}{B \mu}$ (see eq. \eqref{eq:tanbeta}) in terms of  fundamental Lagrangian parameters.
 Once again, we can see that small $\sim 10^{-3}$ variations in $m^2_{H_d}, \mu$ and $B \mu$ cannot change $\tan \beta \sim 10$ appreciably. That is, as we make small detuning of input parameters, the two Higgs doublet VEVs, $v_u, v_d$ change dramatically in size, but their ratio $\tan \beta \equiv v_u/v_d$ does not. Therefore the large stau-mixing is robust against small detuning, and if the weak scale dramatically rises upon detuning, the massive-photon phase transition necessarily occurs.
 
 Turning to the second subtlety, when Higgs condensation is switched off by detuning,  one may wonder if  sneutrino condensation can replace Higgs condensation to achieve an anthropically favorable EWSB with massless photon. But with only small variations of fundamental parameters this too cannot happen. In the region of parameter space we are exploring up to small variations, the {\it only} fine-cancellation is that of the Higgs in  
 eq.~\eqref{eq:EWSB}, to create a much smaller weak scale than the soft masses and $\mu$. In particular, the sneutrino soft masses (resulting from gaugino-mediation) are {\it not} fine-tuned to be at the weak scale, but given by order one factors multiplied by the wino/bino masses. Small variations in these wino/bino masses therefore leads to only small variations in the robustly non-tachyonic sneutrino masses. See eq. \eqref{eq:gauginomedsquarkmassall}. This is the distinction between Higgs and sneutrino condensations: the (lighter) Higgs doublet mass-squared is a difference of two large soft terms where small changes can move us between positive or negative Higgs mass-squared, whereas in gaugino-mediation the sneutrino mass-squared is always positive.

We briefly comment on  anthropic considerations in similar spirit in other scenarios. 
In Split \cite{Wells:2003tf, Wells:2004di, Arkani-Hamed:2004ymt, Giudice:2004tc, Arkani-Hamed:2004zhs} and mini-split SUSY \cite{Arvanitaki:2012ps, Arkani-Hamed:2012fhg} one appeals to approximate $R$-symmetry to explain the hierarchical spectrum of gauginos much lighter than sfermions, and identifying the LSP as thermal WIMP DM to motivate the gauginos $\sim$ TeV.  In RPV Split-SUSY, it was shown \cite{Cui:2013bta} that a WIMP baryogenesis mechanism can be realized (with DM then arising from some other disconnected sector of particle physics), but with
weak-scale gauginos it was necessary for scalars $> 100$ TeV in order to ensure baryogenesis is successful. In each of these scenarios, there is the possibility that the LHP is arising because of some ``compromise'' between naturalness and anthropic selection among a large set of MSSM-like effective theories realized in a multiverse of cosmological spacetime patches. And in each case, there is a striking consequence if one occupies a more natural part of the parameter space of the SUSY theory. In (mini)-Split SUSY if the scalars are brought significantly closer to the weak scale,  the approximate $R$-symmetry then implies very light gauginos, resulting in an  extremely small DM abundance. In the RPV MSSM the consequences are even more dramatic: if the gauginos remains at the weak scale and scalars are also brought down significantly to the weak scale to make the theory more natural, the baryogenesis mechanism breaks down and furthermore any pre-existing baryon asymmetry from a higher scale baryogenesis mechanism is washed out, so that the universe contains very few baryons. Plausibly, anthropic selection favors the stronger production of DM (or baryon asymmetry in the RPV case) due to its role in galaxy formation (or as constituents of complex lifeforms), although such selection is notoriously hard to formulate quantitatively. 
 Nevertheless, these models at least illustrate that being natural and being ``hospitable to intelligent life'' may conflict, and that this may lie at the root of the observed LHP as a compromise between naturalness and anthropic ``pressures''.  In the case of Sleptonic SUSY the anthropic dangers are arguably more clear-cut. 
 There are other frameworks in the literature in which there are dramatic transitions in the BSM physics that are very sensitive to the weak scale, which then can strongly affect the naturalness considerations \cite{Graham:2015cka, Arvanitaki:2016xds,Csaki:2020zqz,  Arkani-Hamed:2020yna, TitoDAgnolo:2021nhd, TitoDAgnolo:2021pjo}. However, each of these has a somewhat different  character from the sense of ``living dangerously'' in Sleptonic SUSY.
 
 Given that in Sleptonic SUSY, living dangerously arises from the hierarchical spectrum, one can ask why in a multiverse of MSSM realizations one simply did not live with considerably less hierarchical soft terms, in which case the phase transition to Higgsed EM would not arise to compromise naturalness. We cannot completely answer such a question, even within the higher-dimensional UV EFT in \autoref{sec:6D}.
  Only the vanishing of the UV sfermion soft terms is explained due to the extra-dimensional sequestering structure.  The remaining UV soft terms are still continuous input parameters at this level, and ultimately would require a more fundamental UV completion, at the level of a more complete field theory or string theory, to understand the {\it necessity} of their hierarchical structure. 
  
  We give a simple example of how {\it some} of the hierarchical structure underlying Sleptonic SUSY, namely the UV gaugino hierarchy, can be sharply determined by an even more UV-complete description of their origin. The simplest such example will yield a Sleptonic SUSY spectrum of a particular type, somewhat different than the benchmarks we later pursue, but nevertheless illustrating the general point regarding hierarchical structure being potentially ``locked in'' by the far UV. We take these gaugino masses (and their associated tree couplings to $X$) to vanish at tree-level, but then get a one-loop gauge-mediated contribution \cite{ Giudice:1998bp} from a vector-like ``messenger multiplet''  living on the SUSY-breaking boundary in \autoref{figure:gauginomed6D}, which themselves feel SUSY-breaking by coupling to $X$ in their superpotential.\footnote{We can think of $X$ as carrying Peccei-Quinn (PQ) symmetry under which $H_u$ and $H_d$ have the same charge so that the $\mu$ term can arise from the PQ-symmetric Kahler coupling  $ \propto X^{\dagger} H_u H_d$. If the messenger multiplets carry analogous charges, they can PQ-symmetrcially couple to $X$ the same way. But then, PQ symmetry would forbid gaugino masses at tree level.} The gauge-mediated gaugino masses induced are then of order $ \sim \frac{\alpha}{4 \pi}F_X/M_{\rm messenger}$, which matches the previous $\sim F_X/M$ of eq.~\eqref{eq:SUSYbreakingLag} if we take $M_{\rm messenger} \sim \frac{\alpha}{4 \pi} M$. Note that relative to  standard gauge-mediation in a purely 4D context the sfermion mass-squareds are suppressed in the UV by $1/(M_{\rm messenger} L)^2 \ll 1$ \cite{Mirabelli:1997aj}, again approximately matching the vanishing of UV sfermion masses as before. 
   
   In this sense, the messenger module we have added is a partial UV-completion of the soft terms in the higher-dimensional set-up. Furthermore it can be more predictive because the detailed gaugino masses depend on the discrete gauge representation of the messenger multiplet. If we take the KK scale close to the unification scale  $\sim 10^{15} - 10^{16}$ GeV then all gauge couplings are nearly the same there, and gaugino mass ratios in the UV are determined entirely by the gauge quantum numbers of the messengers.\footnote{We take hypercharge to be quantized as in its unification into $SU(5)$, so that the gauge quantum numbers of the messenger is a discrete choice.}
       For example, if we choose the vectorlike messenger multiplet to have the same gauge quantum numbers of a SM left-handed quark doublet (and its conjugate),  the ratios of UV gaugino masses is determined to be $M_3: M_2: M_1 = 1:\frac{3}{2}:\frac{1}{10}$ in the UV, which after running into the IR yields $M_3: M_2: M_1 = 1:0.55:0.02$.
Through gaugino-mediation this will result in light bino and right-handed sleptons, with heavier wino, left-handed sleptons, and even heavier colored superpartners. If the stops are $\sim 10$ TeV, the bino and right-handed sleptons will be around the weak scale, realizing one possible spectrum of Sleptonic SUSY type.

 In general, far-UV structure may be difficult to know in detail, in particular if it involves string states. Therefore it is hard to assess the ``living dangerously'' criterion by comparing large, even qualitative, changes to the parameters of the theory without knowing the full UV structure and its prevalence in the multiverse. Instead, we may practically be limited to the ``local'' test to see if small changes of parameters to make the theory more natural trigger (anthropically dangerous) phase transitions. It is notable that Sleptonic SUSY at least passes this local test.

\section{Higher-dimensional Framework incorporating Unification} \label{sec:6D}

As reviewed in \autoref{sec:gauginomediation}, gaugino-mediation requires an extra dimensional interval  for sequestering, in which the three generations of matter are localized on one boundary while SUSY-breaking is localized on the other. 
 On the other hand, the simplest models of SUSY unification, ``Orbifold GUTs'' (for a review, see, for example, \cite{Hall:2002ea}
 and reference therein), also employ an extra-dimensional interval and the mechanism of Split Multiplets, 
 with SM matter now in the bulk so as to allow non-unification of Yukawa couplings (in the lighter generations) and suppressed dimension-$6$ proton decay, while still explaining the attractive quantization of hypercharge offered by traditional GUTs. The Higgs multiplets are also in the bulk to straightforwardly solve their doublet-triplet splitting problem. 
  To combine both the sequestering and unification considerations, one is then led to  consider {\em two} extra dimensions \cite{Haba:2002ve, Buchmuller:2005ma}.
  
  \subsection{$6D$ Geography}
  
\begin{figure}[h]
\centering
\includegraphics[width=0.55 \linewidth]{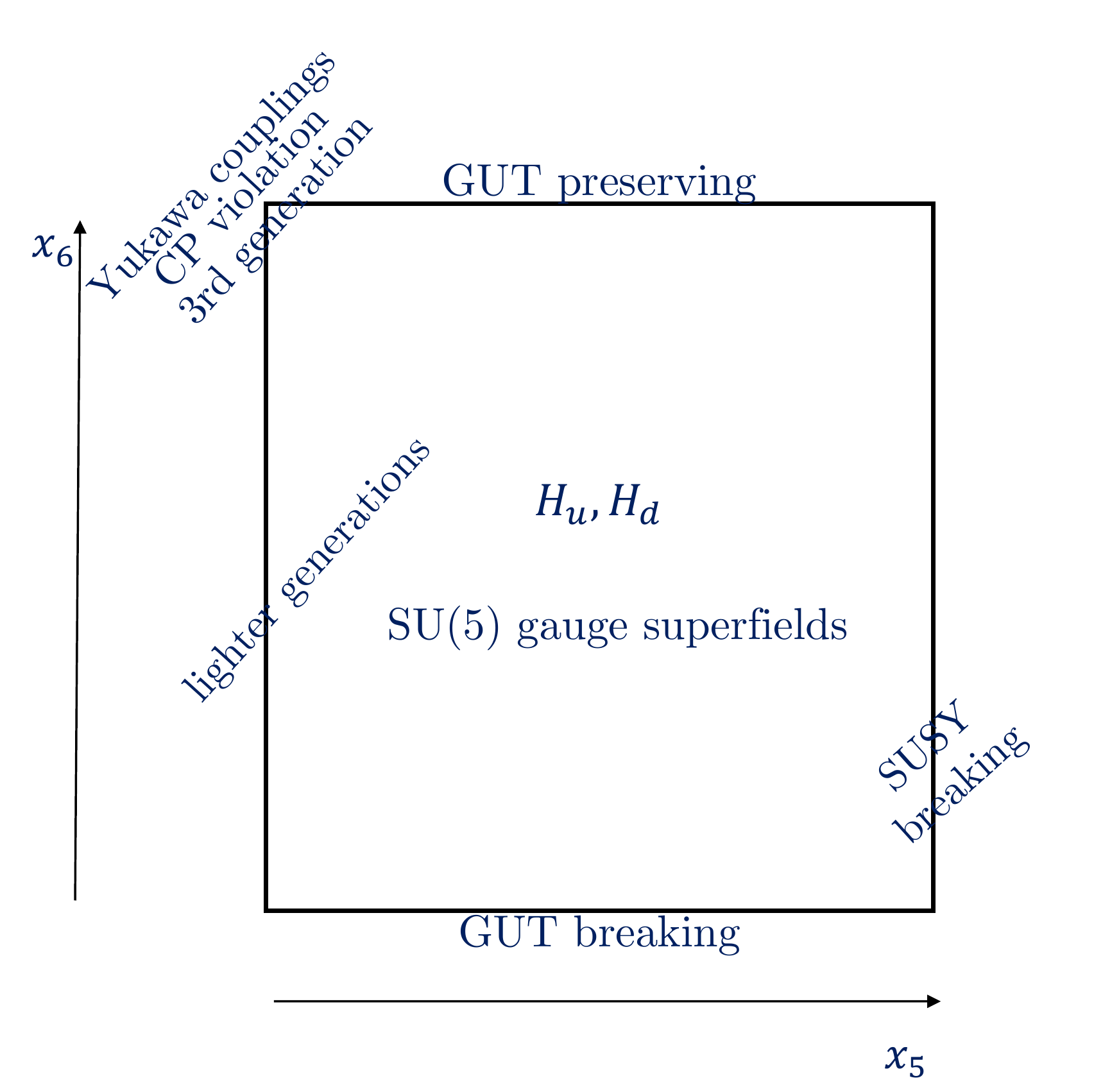}
\caption{Schematic picture of the six-dimensional structure incorporating unification, where only the two extra dimensions are shown. The $x_5$ direction enforces sequestering the SUSY breaking from SM matter fields, flavor-violation and CP-violation. The $x_6$ direction realizes orbifold grand-unification, its Kaluza-Klein scale setting the scale of gauge-coupling and $b-\tau$ unification.} 
\label{figure:gauginomed6D}
\end{figure}

  The extra-dimensional space is  simply  depicted in \autoref{figure:gauginomed6D}. SM matter is localized on the LHS and SUSY-breaking on the RHS, enforcing sequestering so that SUSY-breaking cannot introduce new flavor violation, whereas the Higgs and unified gauge superfields propagate in the bulk so they 
  can acquire tree-level SUSY breaking. 
  With CP only broken on the LHS, there can be a CKM phase for observed CP violation, but no new CP violation beyond that from SUSY-breaking. All this horizontal localization and sequestering is essentially the same as in \autoref{sec:gauginomediation}. We take the unified gauge group to be the canonical choice, $SU(5)$. The vertical localization and boundary conditions (b.c.) then enforce the breaking of the GUT down to the SM. We have a  GUT-preserving b.c. on the top edge and GUT-breaking b.c. on the bottom edge. The mechanism of Split Multiplets can then be realized, with the features mentioned above. We take the third generation to be localized on the top left corner, so that it is sequestered {\it and} its Yukawa couplings are unified, allowing us to retain the attractive feature of $b-\tau$ unification. The lighter generations are localized to the left boundary but propagate to the bottom GUT-breaking edge, and with split multiplets they do not have unified Yukawa couplings, compatible with observations. Furthermore, if Yukawa couplings to the Higgs fields are themselves localized to the top left corner, the first two generation Yukawa couplings will have an extra suppression relative to the third generation in the 4D EFT, due to the ``dilution'' in  the vertical dimension, partially explaining why the heavier generation is the one that has unified Yukawa couplings. 
  
 In order for 
  precision gauge coupling unification to proceed via the 4D SUSY RGE, it is important that there are no KK scales below the unification scale,  this being the KK scale associated to GUT breaking in Orbifold GUTs. It follows that 
   the sequestering KK scale must be comparable or higher than this unification scale  (though of course lower than the Planck scale). That is, $L_{\rm GUT} \geq L_{\rm sequester}$, explaining in more detail why we are necessarily in 
   a high-scale SUSY breaking scenario, so that we must carefully track the large RG evolution effects to $\sim$ TeV scales as we have been doing. 

Beyond the general requirement that SUSY breaking be localized in the sequestering dimension on the RHS, it can further be localized in the GUT breaking dimension or not, yielding two robust options. If not localized or localized at the top right corner, then we obtain the familiar pattern of (nearly) unified gaugino masses at the KK (unification) scale. As mentioned in \autoref{sec:OneLoopRGE}, while significant RG effects will tend to drive the gauginos apart in the IR, this effect is insufficient to realize the Sleptonic SUSY hierarchy. Instead we localize the SUSY breaking 
 at the bottom right so that the gaugino masses are not unified already at the gauge-coupling unification scale, plausibly with modest hierarchy $M_3 > M_2 > M_1$.
 It is this latter robust option that can lead in the IR to Sleptonic SUSY, with RG effects accentuating the UV gaugino hierarchy, 
 the heavy gluino dragging up squarks to be comparably heavy, while the lighter wino and bino being compatible with light sleptons.

\subsection{Higher-dimensional Non-renormalizability and Sequestering}

The 6D unified gauge theory is a non-renormalizable EFT, and therefore has a maximal energy scale of validity by which new states must appear as part of the UV completion of this EFT. In general, such new bulk states can mediate SUSY-breaking from the SUSY-breaking boundary (right) to the matter boundary (left), giving rise to new flavor/CP-violating SUSY-breaking effects which would threaten the generalized GIM mechanism. But such effects will be Yukawa-suppressed $\sim e^{-M_{\rm new} L_{\rm sequester}}$. In order for this suppression to be consistent with current stringent flavor/CP bounds, without any further structure, we must have 
$\sim e^{-M_{\rm new} L_{\rm sequester}} 
<  10^{-3} \,$ \footnote{The bound on ``anarchic'' squark masses is $\sim 5000$ TeV, whereas we have $\sim 10$ TeV, so we need a suppression of $\sim 10^{-6}$ in the amplitude for
$K-\bar{K}$ mixing. However, this $\Delta S = 2$ amplitude $\propto$ {\em two} powers of $\Delta S = 1$ {\em off}-diagonal squark mass$^2$.}. We can bound $M_{new}$ by the scale at which the 6D gauge theory becomes strongly-coupled in the UV \cite{Chacko:1999hg},  $M^2_{new} < 
32 \pi^2 /\left( N_{\rm GUT} \alpha_{\rm GUT} N_{\rm hel } L_{\rm seq} L_{\rm GUT} \right) $. 
The factor of 
$ 32 \pi^2 / \left( N_{\rm GUT} \alpha_{\rm GUT} N_{\rm hel } L_{\rm seq} L_{\rm GUT} \right)$
is the standard naive dimensional analysis for the strong coupling scale of the 6D gauge coupling that matches the 4D unified coupling of the MSSM: $1/g_{\rm 6D}^2 = L_{\rm seq} L_{\rm GUT}/g_{\rm 4D}^2$, with $\alpha_{\rm GUT} \sim 1/25$. The extra factor of $N_{\rm GUT} = 5$ is due to the large multiplicity of GUT ``colors'' that enhance gauge loops; similarly $N_{\rm hel } \sim 4$  counts number of degrees of freedom/helicities of gauge bosons (where we are being crude but conservative in the estimate here). 
Clearly, the best case for sequestering is that $L_{\rm seq} \sim L_{\rm GUT}$, so that the Yukawa suppression $e^{-M_{\rm new} L_{\rm sequester}} > 10^{-8}$, which is significantly smaller than what is needed, so we are safe. From now on, we will take this optimal sequestering case of comparable KK scales to hold, $L_{\rm seq} \sim L_{\rm GUT}$.

\section{Important two-loop RG Effects} \label{sec:twoloopRGE}

In addition to the one loop RG effects controlled by the gauge and Yukawa couplings that we have discussed so far, there are two-loop contributions which can give important corrections \cite{Martin:1993zk}. At the two-loop order, the heavy scalar masses can feed into the light scalar masses through gauge and Yukawa couplings, with the heavy scale compensating the two-loop suppression. We therefore will keep such two-loop effects. In particular the large squark and Higgs masses feed into the masses of the lighter sleptons as can be seen from the following terms
\bea \label{eq:twoloopEWRGE}
\left( 16\pi^2\right)^2 \frac{d m^2_{L}}{dt}  \supset && \frac{3}{5}g_1^2  \sigma_1 + 3 g_2^2 \sigma_2-\frac{6}{5} S', \\
\left( 16\pi^2\right)^2 \frac{d m^2_{\bar{e}}}{dt} \supset && \frac{12}{5}g_1^2  \sigma_1+\frac{12}{5} g_1^2 S', \nonumber
\eea
where $\sigma_{1,2}$ are defined as
\bea
\sigma_1 && = \frac{6}{5} g_1^2  {\rm Tr}\left[ Y_j^2 m^2_{\phi_j}\right], \\
\sigma_2 && = g_2^2 \left( m^2_{H_u} + m^2_{H_d}+3 m^2_{Q}+m^2_{L}\right), \nonumber
\eea
and $S'$ is given by
\bea
S^\prime \approx && |y_t|^2 \left( -3 m^2_{H_u}-m^2_{Q_3} +4 m^2_{\bar{u}_3} \right)+|y_b|^2 \left( 3 m^2_{H_d}-m^2_{Q_3}-2 m^2_{\bar{d}_3} \right)+|y_\tau|^2  m^2_{H_d} \nonumber \\
  && +\left( \frac{3}{2}g_2^2+ \frac{3}{10}g_1^2 \right) \left(m^2_{H_u}-m^2_{H_d} \right)+\left( \frac{8}{3}g_3^2+\frac{3}{2}g_2^2+ \frac{1}{30}g_1^2 \right) {\rm Tr}(m^2_{Q}) \nonumber \\
  && - \left( \frac{16}{3}g_3^2+\frac{16}{15}g_1^2 \right) {\rm Tr} (m^2_{\bar{u}}) +\left( \frac{8}{3}g_3^2+\frac{2}{15}g_1^2 \right) {\rm Tr} (m^2_{\bar{d}}).
\eea
These contributions to soft slepton masses are flavor-universal. We now comment on various terms above starting from 
the contributions given by $\sigma_{1,2}$. If these terms are dominated by the positive squark mass-squareds (that is if Higgs soft masses are not too large), or if $m^2_{H_{u,d}}>0$, they result in negative contributions to the slepton mass squareds. In this regime, their effect should be kept subdominant to the gaugino mediation contribution to avoid tachyonic (or too light) sleptons. For this to be the case, the hierarchy between slepton and squark masses should not be too large. This condition can be written roughly as
\beq
\frac{m^2_{\rm squark}}{m^2_{\rm slepton}} \lesssim \frac{(16 \pi^2)^2}{\log (M_{\rm GUT}/{\rm TeV})} \sim (30)^2.
\eeq
There is another regime, where negative and large $m^2_{H_{u,d}}$ dominate $\sigma_{1,2}$ and leads to an increase in slepton masses. This regime was used by \cite{Cox:2018vsv}
(see also refs.~\cite{Yamaguchi:2016oqz,Yin:2016shg, Yanagida:2018eho, Yanagida:2020jzy} for previous
applications of such negative $m^2_{H_{u,d}}$)
  to raise the slepton masses and make bino the LSP, instead of using the D-mediation effect discussed in \autoref{subsec:Dmediation}. This however works only for very large $\tan \beta \sim 50$, since with moderate $\tan \beta \sim 10$ and starting with very large and negative $m^2_{H_u}\approx m^2_{H_d}$ in the UV, the RG running makes $m^2_{H_d}<m^2_{H_u}$ in the IR. This would then result in large $H_d$ condensation and hence not in the desired EW symmetry breaking.

While the terms given by $\sigma_{1}$ contribute with the same sign to the doublet and singlet sleptons, the contribution from $S'$ is proportional to the hypercharge 
which has opposite signs for the doublet and singlet sleptons.
In this regard, it is similar to the contribution of eqs.~\eqref{eq:xiinslptonmassLR}, and therefore can be counter-balanced by adjusting $\xi$. 

Ordinarily, when one solves RG evolution from the GUT scale at two-loop precision, one must consistently use UV boundary conditions given by one-loop matching at the GUT scale. But in the present context, we are not using two-loop RG effects in order to get high precision, but rather to get important RG effects that feed from the heavy soft masses of colored and Higgs fields down to the uncolored light fields, which are comparable to the one-loop RG
effects of the light fields among themselves. For this restricted purpose, we can continue to use tree-level matching at the GUT/sequestering (KK) scale. 
With these tree-level UV conditions, and 
solving the RGEs numerically, we obtain following expressions for the slepton soft mass squareds :
\bea
&&m^2_{L_2} \approx (0.44)^2 M_1^2 + 0.8^2 M_2^2 - (0.32)^2 \left(\frac{M_3}{10}\right)^2 + \left[- 0.4^2+(1.6)^2 \xi\right] \left(\frac{m^2_{H_d}}{10^2}\right),
  \label{eq:MLMRRGE2loop} \\ 
&&m^2_{\bar{e}_2} \approx  (0.88)^2 M_1^2- (0.29)^2 \left(\frac{M_3}{10}\right)^2+\left[ (0.24)^2-(2.2)^2 \xi\right] \left(\frac{m^2_{H_d}}{10^2}\right),  \nonumber \\
   \label{eq:MLMRRGLsplit2loop}
&&m^2_{L_3}-m^2_{L_2} \approx  (0.03)^2 \left(\frac{M_3}{10}\right)^2 \left(\frac{\tan \beta}{10} \right)^4\\ \nonumber 
&&- \left[  (0.05)^2 M_1^2 + 0.05^2 M_2^2 + (0.52)^2 \left(\frac{m^2_{H_d}}{10^2}\right)\right]\left(\frac{\tan \beta}{10} \right)^2,   \\ \nonumber
&&m^2_{\bar{e}_3}-m^2_{\bar{e}_2} \approx  (0.04)^2 \left(\frac{M_3}{10}\right)^2 \left(\frac{\tan \beta}{10} \right)^4\\ \nonumber
&&- \left[ (0.07)^2 M_1^2+0.07^2 M_2^2+ (0.74)^2\left(\frac{m^2_{H_d}}{10^2}\right)\right]\left(\frac{\tan \beta}{10} \right)^2, 
\eea
where all the soft SUSY breaking parameters are evaluated in the IR. Equations \eqref{eq:MLMRRGE2loop} give the universal contributions to slepton masses. In these two equations, the terms proportional to $M_{1,2}^2$ are the one loop gauge mediation contributions and the terms proportional to $\xi$ are the D-mediation contributions discussed in \autoref{sec:OneLoopRGE}. The remaining terms, which are proportional to $M_3^2$ and $m^2_{H_d}$, arise from the two loop corrections discussed in this section.  
Finally equations \eqref{eq:MLMRRGLsplit2loop} give the splitting between the third generation and other sleptons controlled by the $\tau$ Yukawa coupling. In the region of the parameter space with  large soft Higgs masses, the leading effect is given by terms proportional to $m^2_{H_d}$ which are the one-loop contributions of eqs. \eqref{eq:RGEHiggsforL3E3}.

\section{Vacuum (Meta-)stability and a Non-minimal Model Construction} \label{sec:stability}

In this section, we discuss the bounds on Sleptonic SUSY coming from vacuum stability considerations. First we will consider the condition that the standard vacuum be a local minimum of the potential. Then we will see that even if this condition is satisfied, there may still be other more stable minima to which the standard vacuum can tunnel, giving rise to stronger cosmological bounds. We will also present a non-minimal model in which in some examples the EW vacuum can be absolutely stable, while in others it can be cosmologically stable but displaying a broader range of phenomenologically interesting features than in the MSSM.

\subsection{Perturbative Stability}

We focus on the potential for the lightest scalars in Sleptonic SUSY. These are the lighter Higgs (which is predominantly $H_u$) and the sleptons. For selectron and smuon, because of absence of large trilinear couplings (which are proportional to corresponding Yukawa couplings), it is sufficient to require non-tachyonic mass terms. So in the following discussion we will assume that is the case and will set the fields for smuons and selectrons as well as sneutrinos to zero. 
The terms of the potential involving the neutral component of $H_u$ and the staus is given by
\bea \label{eq:thepotential}
V=& (m^2_{Hu}+\mu^2)|H_u^0|^2+m^2_{L_3}|\tilde{\tau}_L|^2+m^2_{\bar{e}_3}|\tilde{\tau}_R|^2  - (y_\tau \mu H_u^0 \tilde{\tau}_L^* \tilde{\tau}_R+\rm{h.c.}) +y_\tau^2 |\tilde{\tau}_L \tilde{\tau}_R|^2 \\ \nonumber 
&  +\frac{g_2^2}{8}(|\tilde{\tau}_L|^2+|H_u^0|^2)^2 
 +\frac{g_Y^2}{8}(|\tilde{\tau}_L|^2-2|\tilde{\tau}_R|^2-|H_u^0|^2)^2+\delta \lambda_H |H_u^0|^4
\eea
 where $\tilde{\tau}_{L,R}$ denote the staus. The mass term for $H_u$, i.e. $(m^2_{H_u}+\mu^2)$, and radiatively-generated quartic $\delta \lambda_H$ are fixed such that they give the observed $125$ GeV Higgs mass and the correct EW breaking VEV for the Higgs. Note in particular the presence of a trilinear coupling proportional to $\mu$ and $y_\tau \approx \frac{\tan\beta}{100}$.
 This coupling  plays an important role in the vacuum stability considerations.

  By setting $H_u$ to its SM VEV, the cubic coupling of eq.~\eqref{eq:thepotential} turns into a mixing term for staus. If this mixing is too large, it leads to a tachyonic stau mass eigenstate. The mass-squared matrix for the staus is given by
 \beq
 M^2_{\tilde{\tau}}\approx
 \begin{pmatrix}
 m^2_{L_3} & - v \mu y_\tau\\
  - v \mu y_\tau & m^2_{\bar{e}_3}
\end{pmatrix}
 \eeq
In order for the standard vacuum to be a minimum (and not a saddle point) of the potential, the above mass squared matrix should have only positive eigenstates, leading to the following  condition:
\beq
 m^2_{L_3} m^2_{\bar{e}_3}> ( v \mu y_\tau)^2 \approx ( 0.4 \, {\rm TeV})^4 \left(\frac{\mu \tan \beta}{100 \TeV}  \right)^2.
\eeq

\subsection{Non-perturbative Stability}
Even if the stau mass squared matrix does not have negative eigenvalues, and the standard EW vacuum is locally stable, there may still be deeper minima in the potential of eq.~\eqref{eq:thepotential}, to which the standard vacuum can tunnel \cite{Rattazzi:1996fb,Hisano:2010re,Carena:2012mw,Kitahara:2013lfa}.
 This is because of the presence of the trilinear term in the potential $\propto y_\tau \mu$. If present, these minima lie in field regions where $\langle H_u^0 \rangle, \langle \tilde{\tau}_R \rangle, \langle \tilde{\tau}_L \rangle \neq 0$, spontaneously breaking electromagnetism. The other minima may however also have other scalar VEVs beyond just these. 
 
\begin{figure}[h] 
\centering
\includegraphics[width=0.7 \linewidth]{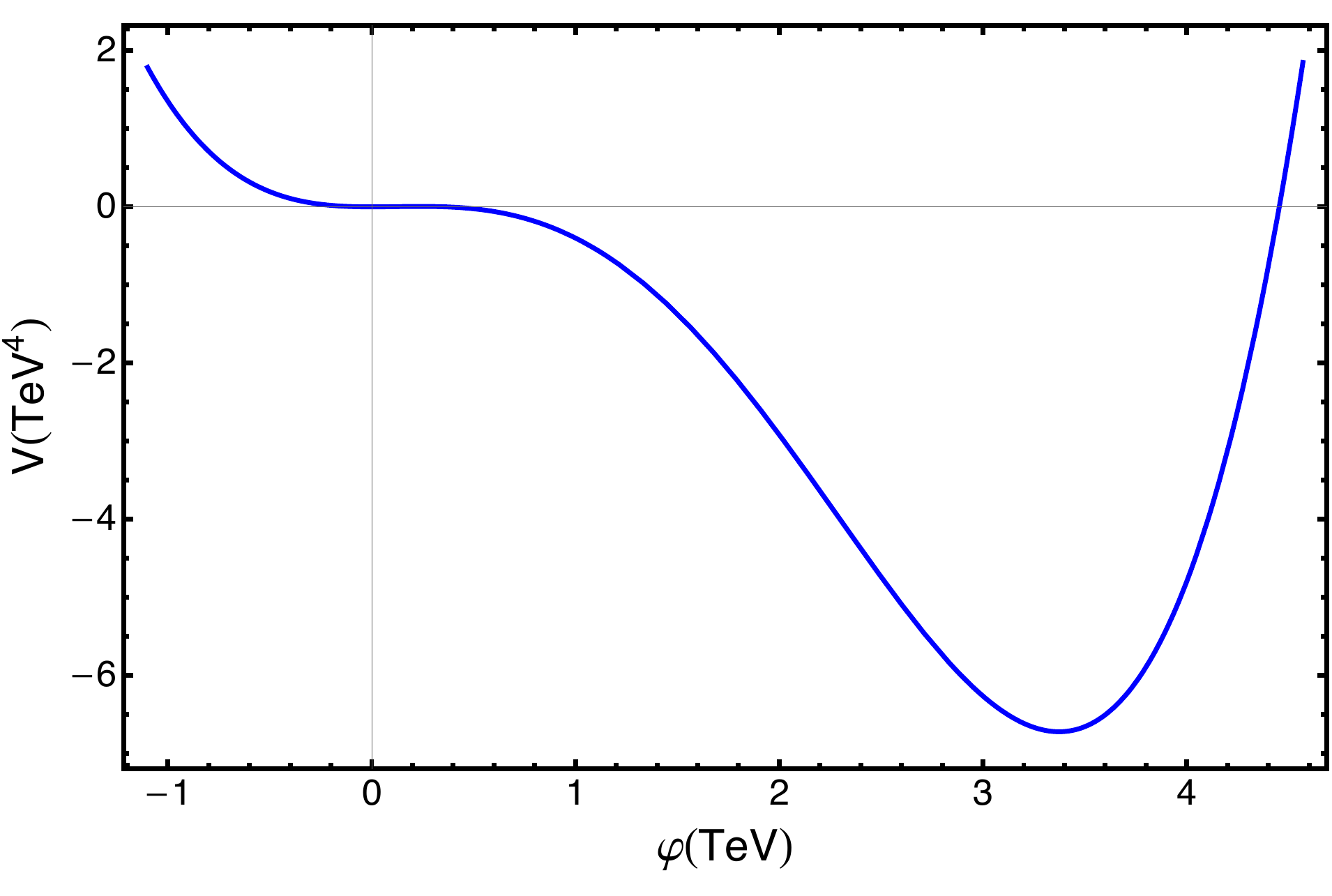}
\caption{The potential along the field direction $H_u^0=v+\varphi/\sqrt{2}$, $\tilde{\tau}_L=1.4 \frac{\varphi}{\sqrt{2}}, \tilde{\tau}_R=1.5 \frac{\varphi}{\sqrt{2}}$ and for the parameters of the first benchmark in \autoref{tab:bmminimal}. The standard EWSB vacuum ($\varphi=0$), is a local minimum of the potential and meta-stable. But there is a much deeper minimum at larger field values due to the sizeable stau-Higgs trilinear scalar coupling that accompanies regimes in which there is a sizeable muon $g-2$ contribution. \label{fig:EMbreakingminimum}} 
\end{figure}

 To get some instinct, it is useful to make the simplifying (but ultimately false) assumption that the only non-zero VEVs are $\langle H_u^0 \rangle, \langle \tilde{\tau}_R \rangle, \langle \tilde{\tau}_L \rangle$. With this assumption, we can continue to use the potential of eq.~\eqref{eq:thepotential}. 
  \autoref{fig:EMbreakingminimum} shows this potential along the direction connecting the standard vacuum to one of its charge breaking minima. For concreteness we have chosen the specific parameters that result in the spectrum to be presented in our first benchmark in \autoref{tab:bmminimal}, which is consistent with a BSM correction to the muon $g-2$ comparable to the current discrepancy.

In the presence of this EM-breaking vacuum,  assuming that the cosmological evolution populates the standard/ordinary vacuum, we should then ask if it is long-lived enough to have survived the decay by vacuum tunneling to the EM-breaking vacuum.  

When both vacua are local minima of the potential, this decay happens by nucleation of EM breaking bubbles in the EM preserving background.
Bubble nucleation rate, defined as the probability of bubble nucleation per unit volume per unit time can be written in the form \cite{Coleman:1977py}
\beq
\Gamma\sim M^4 e^{-B} 
\eeq
where the prefactor $M^4$ is set by the mass scales in the potential which are of order $M\sim$ TeV. However the analysis here is not sensitive to its precise value, given the much higher sensitivity to the exponent $B$. $B$ can be found semiclassicaly by computing the Euclidean action of the bounce solution, which is a nontrivial $O(4)$-symmetric solution of Euclidean equations of motion which asymptotes to the standard vacuum at infinity and is regular at the origin/center.
In order for the standard vacuum to have a lifetime longer than the age of the universe, $\Gamma \lesssim H_0^4$,
we should have $B > 4 \ln{ ( \frac{m_{\tilde{\tau}} }{ H_0} ) }\approx 400$. 

We numerically check the above condition for our benchmarks. To compute the bounce action, we use a set of ansatze along the field directions with fixed ratios among $h$,$\tilde{\tau}_L$, and $\tilde{\tau}_R$ fields, solve the equations of motion that are obtained along these directions and then extremize the action of these solutions as the direction is varied. As a test of these fixed-ratio ansatze, we find that they are consistent with the bounds  given by the fitting formulas in \cite{Hisano:2010re} as well as \cite{Kitahara:2013lfa}. 

However, we must now address the possibility of minima with non-zero VEVs beyond those of $H_u$ and the staus \cite{Komatsu:1988mt,Casas:1995pd,Rattazzi:1995gk,Strumia:1996pr}. 
The field regions where these minima may lie are discussed in 
\autoref{app:othervacua}.
To estimate the tunneling rates to these minima, we proceed in the analogous manner to the above discussion, but enlarging our ansatze to cover all the relevant field directions \cite{Casas:1995pd}, with fixed ratios again. We will use  this procedure to provide plausible crude lifetime bounds for the EW vacuum, but the enlarged field space does not yet have corroborating studies along the lines of  \cite{Hisano:2010re, Kitahara:2013lfa}). For our benchmarks with metastable EW vacuum, we found that the bounce actions in the enlarged field space were not lower (EW vacuum lifetimes were larger) 

than those of the simplified stau-$H_u$ field space, so that the simplified analysis is effectively valid for us. 

Reference \cite{Duan:2018cgb} has found that considering the thermal corrections to the potential, the thermal transition from the EW symmetric vacuum to the EM-breaking vacuum may happen at temperatures $\gtrsim O(100)$ GeV, even in parts of the parameter space where the standard vacuum would be meta-stable at late times. This could lead to a stronger constraint than the zero-temperature lifetime bound discussed above. The bound from thermal transition is however more model-dependent as it requires that the universe has been reheated to such high temperatures. We leave for future study the exploration of such subtle cosmological constraints in the specific context of Sleptonic SUSY.  Provisionally, we have just focused for now on satisfying cosmological stability at low temperatures, presuming that the EW vacuum has survived until the universe has cooled to low temperatures. 

Clearly, it would be reassuring to find models in which the subtle issues of cosmological population of a meta-stable EW vacuum~\cite{Strumia:1996pr,Ellis:2008mc}, and the uncertainties involved in calculating its thermal or zero-temperature lifetime, are avoided. We can find such a benchmark (e.g., BM2 in \autoref{tab:bmminimal}) in the MSSM itself by taking small enough $\mu$ that the EW vacuum is absolutely stable, but at the cost of having only very small corrections to the muon $g-2$. 
This motivates us to find a new model in which the spectrum is consistent with both EW vacuum absolute stability as well as a sizeable muon $g-2$ correction.

\subsection{Non-minimal Model}
\label{sec:nonminimalmodel}

In this subsection we introduce a simple non-minimal model/generalization with an extra pair of Higgs doublets (and their color-triplet unification partners) in which it is possible to find parameters for which the EW vacuum is absolutely stable, while still accounting for the 
 muon $g-2$ anomaly.
 Even beyond absolute stability, the non-minimal model allows us to explore an interesting regime of the collider phenomenology which is difficult to realize in the MSSM, namely the case in which the co-annihilation NLSP partner of the DM bino-LSP is a stau. As discussed in \autoref{sec:OneLoopRGE}, this allows the smuon and selectron to be significantly heavier than the Bino LSP and therefore produce significantly more visible hard leptons 
 from 
 their production, followed by decay into the LSP. 
 In the MSSM, the EW vacuum of this light-stau regime typically would have shorter lifetime than the age of the universe. In the non-minimal model, in this regime, the EW vacuum  is not absolutely stable, but can readily be cosmologically stable enough. The non-minimal model also allows us to demonstrate that it is possible for Sleptonic SUSY to give both an observable correction to the muon $g-2$ while also giving an observable signal in dark matter detection.

To understand the basis for our non-minimal construction, we must first appreciate the multiple roles that the $\mu$ parameter plays in the minimal model which strongly constrains the acceptable portions of parameter space. The non-minimal model distributes these roles among separate parameters, considerably freeing the constraints. 
In the minimal model, the same large $\mu$ enters in (fine-)tuning EWSB, enhances  the $g-2$ contribution, and at the same time multiplies the large stau-Higgs trilinear couplings which is responsible for the instability of the standard vacuum. 
Furthermore, as negative $m^2_{H_d}$ is used to raise the stau masses, it is $\mu$ that prevents large $H_d$ condensation. We first show that in a non-minimal model these roles can be played by different parameters, allowing for an explanation of muon $g-2$ anomaly while maintaining the absolute stability of the standard EWSB vacuum. 

Lower $\mu$, thought of as a Higgsino mass term, also allows greater bino-Higgsino mixing angle after EWSB, which is central to the DM direct detection cross-section. One can therefore ask if there is a middle-ground for $\mu$, where it is large 
enough to give sizeable muon $g-2$ corrections, while small enough to give observable DM direct detection. This requires larger $\tan \beta$ to compensate the non-maximal $\mu$. In the MSSM, this strategy is problematic, because the smaller $\mu$ means that EWSB fine-tuning must be achieved with large {\it positive}
$m_{H}^2$ 
in the UV, 
which lowers the stau mass by Yukawa-mediation, relatively raising the smuon mass which again results in a suppressed $g-2$ contribution. By contrast, 
in our non-minimal model, we do not need very large $m_{H}^2$ 
in the UV, 
because EWSB fine-tuning is done 
instead by a different parameter, denoted by  $\mu_u$, which is large, while
Higgsino mass is kept small. 
We are therefore able to find parameter regions in which significant $g-2$ corrections and direct detection cross-sections co-exist.

Our non-minimal model is based on introducing two new chiral superfields  $H^\prime_u$ and $H^\prime_d$ with the same quantum numbers as $H_u$ and $H_d$ respectively. We will take the $H^\prime$ to have no Yukawa couplings to SM matter fields. This is easily enforced by assuming that in the higher-dimensional UV realization, the $H^\prime$ are localized to the boundary on which SUSY is broken while SM matter is localized elsewhere. With these extra Higgs multiplets, there are now four possible generalized $\mu$-terms in the superpotential,
\beq
W \supset \mu H_u H_d +\mu_d H_u^\prime H_d +\mu_u H_u H_d^\prime +\mu^\prime H_u^\prime H_d^\prime.
\eeq
The different roles played by $\mu$ in the minimal model are now shared among these different $\mu$ terms of the non-minimal model. Relatedly, there can be new generalized $B \mu$ terms. We are taking all such terms to be ``small'' $\propto 1/ \tan \beta$. We are experimentally very sensitive to $\langle H_d \rangle = \langle 
H_u \rangle/ \tan \beta$ because it gives non-zero masses to all down-type SM fermions, but other non-minimal effects of order $1/ \tan \beta$ can be safely neglected.\footnote{ The smallness of all $B \mu$ terms is 
protected by approximate Peccei-Quinn (PQ) symmetry, in which the SUSY-breaking field $X$  carries charge $+2$ and {\it all} Higgs multiplets have charge $+1$, so that all the generalized $\mu$ terms can be symmetrically generated via the Giudice-Masiero mechanism. The generalized $B \mu$ terms then must break PQ symmetry explicitly, and we can take this breaking to be ``small'' in the above sense.}

The scalar mass-squared for $H_u$ is now controlled by the combination $(m^2_{H_u}+|\mu|^2+|\mu_u|^2)$. This allows for the possibility that the (fine-)tuning of EWSB is dominantly between $|\mu_u|^2$ and $m^2_{H_u}$, while $\mu$ can be chosen to be smaller. Similarly, the mass term for $H_d$ becomes proportional to $(m^2_{H_d}+|\mu|^2+|\mu_d|^2)$. It is therefore possible to avoid a tachyonic mass term for $H_d$ scalar (and hence a large $H_d$ condensate) by choosing a large $\mu_d$ rather than $\mu$.
This allows us to have a large non-universal $H_d$-mediated contribution to the stau masses, along the lines of eqs. \eqref{eq:RGEHiggsforL3E3}, without having $H_d$ condensation by taking large enough $\mu_d$ 
(while keeping $\mu$ small), 
thereby making the preferred EWSB vacuum more (and even absolutely) stable. 
In the minimal model, this requires taking large enough $\mu$ which then also increases  the trilinear stau-Higgs coupling which deepens the unwanted EM-breaking vacuum, whereas there is no such correlated shift in coupling in the non-minimal model. In the Appendix, we present a toy scalar field model which correctly captures the parametrics of vacuum (meta-)stability and gives a simple intuition for how the non-minimal model has a more robust EWSB vacuum than the minimal model. For our meta-stable EW vacuum benchmark, we will continue to use the procedure outlined in \autoref{sec:stability} to check its cosmological lifetime.

We now discuss some of the important new effects that arise in the non-minimal model. In particular, the new $H^\prime$ scalars mix with the $H_{u,d}$ scalars and acquire VEVs. There is also a new trilinear couplings between sleptons and $H^\prime_{u}$ that leads to a correction to the muon $g-2$ that needs to be taken into account. Note the following new terms in the scalar potential: 
\begin{equation}
V \supset (\mu_u \mu^{\prime *} +\mu \mu_d^*) H_u^{\prime \dagger} H_u - y_i \mu_d^* H_u^{\prime \dagger} \tilde{L_i} \tilde{\bar{e}}_i+ {\rm h.c.} +(m^2_{H_u^\prime}+|\mu^\prime|^2+|\mu_d|^2) |{H_u}^\prime|^2 \;
\end{equation}
The mixing term between $H_u$ and $H_u^\prime$ leads to a VEV for $H_u^\prime$,
\beq \label{eq:Huprimevev}
\langle H_u^\prime \rangle \sim - \frac{\mu_u \mu'+\mu \mu_d}{m^2_{H_u^\prime}+|\mu^\prime|^2+|\mu_d|^2}  \langle H_u \rangle
\eeq
Given that the approximately SM-like Higgs doublet is identified with $H_u$, while all other Higgs doublets are extremely heavy, EW precision constraints are straightforwardly satisfied except for the possibility of Higgs precision constraints that require that the $H_u$ predominantly contains the $125$ GeV physical scalar and $\langle H_u \rangle$ dominates EWSB. In particular here, with small generalized $B \mu$ terms, this reduces to
 $\frac{\langle H_u^\prime \rangle}{v} \lesssim 0.1$~\cite{Gu:2017ckc,Carena:2018vpt}. 
 This can be achieved by choosing large enough $m^2_{H_u^\prime} \sim {\cal O}((20-30 {\rm TeV})^2)$ \footnote{Note that 
 such $H^\prime$ soft masses do not radiatively destabilize the $\sim 10$ TeV $H$ soft masses.}.
 
 Furthermore, the trilinear coupling, once $H_u^\prime$ is set to its VEV, mixes the doublet and singlet smuons, which contributes to the muon $g-2$, analogously to the contribution  $\propto \mu$ appearing in the MSSM. 
 The net effect is that the BSM muon $g-2$ contribution is corrected in going from the MSSM to the non-minimal model
 as
\beq
\Delta a_\mu \rightarrow \Delta a_\mu \left(1- \frac{\mu_d}{\mu} \frac{\langle H_u^\prime \rangle}{v}\right).
\eeq
 The non-minimal contribution is subdominant for $\mu_d \sim \mu$, although we
 will nevertheless include this correction in the benchmarks presented in the next section.
Similar to the mixing between $H_u$ and $H_u^\prime$, $H_d^\prime$ also mixes with $H_d$ and acquires a VEV $\propto  \langle H_d \rangle\approx v/\tan \beta$, the effects of which can be ignored for large $\tan \beta$.

The new $H^\prime$ fields also imply a new $\xi'$ parameter analogous to $\xi$ for the $H$ fields which will also contribute to the $D$-mediation effects, along the lines discussed in \autoref{sec:nonminimalmodel}. In the fundamental higher-dimensional construction $\xi$ and $\xi'$ are both forms of custodial isospin breaking on the SUSY-breaking boundary. We will therefore take them to be comparable, $\xi' \sim \xi$. Furthermore, we will take $\xi$ to be comparable to the KK threshold correction it receives in passing to the 4D EFT (due to top loops),
$\xi \sim 3 y_t^2/8 \pi^2$.\footnote{In the MSSM case, we have taken 
$\xi > 3 y_t^2/8 \pi^2$, so that the loop-level KK threshold corrections are subdominant.}  Effectively, this allows us to take $\xi \sim \xi' \sim {\rm few} \%$ in the UV of the 4D RGE (without any fine-tuning of the higher-dimensional parameters). Because, the $H'$ soft masses are significantly larger than the $H$ soft masses, $\xi'$ will dominate the $D$-mediation effects in the non-minimal model.

A final remark is that in the non-minimal model, to keep gauge coupling unification, we assume that $H^\prime_{u,d}$ are added with colored partners, so that together they form $5$ and $\bar{5}$ representations of $SU(5)$. Furthermore these color-triplet components should have masses not far from the doublet components.
The presence of the new states in the non-minimal model modifies the RGE effects compared to the MSSM, with two main effects. Firstly, the correction to gauge coupling running leads to larger unified gauge coupling at the unification scale. This enhances the gaugino-mediated masses for the sfermions.  Secondly, the large soft masses for the new states feed into slepton masses via the two-loop effects parallel to the ones discussed in \autoref{sec:twoloopRGE}. We will include all these effects for our benchmarks, which are presented below. 

\section{Benchmark Models} \label{sec:benchmarks}

In this section, we present benchmark (BM) points for the MSSM and our non-minimal model, taking into account the particle phenomenological constraints as well the constraints on the IR parameters that are imposed by the UV structure and subsequent RG evolution. They illustrate a number of distinct qualitative possibilities.
All five of the benchmark spectra are (partially) accessible at the LHC, given dedicated searches and improvements in the (not-so-hard) lepton-rich final states, as well as ample guaranteed discovery capability at future colliders.
For each of our benchmarks, we identified the most sensitive channels covered by current LHC searches, typically given by the multilepton searches for the Winos~\cite{ATLAS:2019wgx,CMS:2021cox,CMS:2022nty}, and checked that our benchmarks are still viable.
We checked and rescaled the current limits with corresponding branching factions and cross sections calculated through {\tt SoftSUSY}~\cite{Allanach:2001kg}.
BM1 -- BM4 conserve $R$-parity so that their bino LSP serves as a DM candidate, with the NLSP-LSP splitting yielding the observed co-annihilating DM relic abundance. BM5 is R-parity violating, with slepton LSP, with either prompt or long-lived decays being viable depending on the strength of the $R$-parity violating coupling. Our benchmarks illustrate how the potential muon $g-2$ anomaly might be accounted for by Sleptonic SUSY, but also illustrate how the SUSY corrections can straightforwardly be very small, consistent with possibility of the anomaly disappearing  with improvements in SM theory calculations.

BM1 gives a MSSM example in which the  muon $g-2$ SUSY correction is approximately the size of the  potential anomaly, $\Delta a_{\mu} \approx 
2.5\times 10^{-9}$.  But the (spin-independent) DM-proton direct detection cross-section,  $\sigma_{\rm SI} \sim \mathcal{O} (10^{-53}) {\rm cm}^2$, is far below the sensitivity of planned experiments.

BM2 illustrates the MSSM trade-off between getting a sizeable 
$\Delta a_{\mu}$ comparable to the potential anomaly and getting an observable direct detection DM cross-section. In BM2, $\Delta a_{\mu} \ll
10^{-9}$, but now the spin-independent DM-proton cross-section $\sigma_{\rm SI} \simeq 2\times 10^{-47} {\rm cm}^2$ is within reach of the next generation of DM experiments. 

BM3 illustrates that {\it absolute} stability of the EW vacuum is readily achieved within our non-minimal model while still achieving a sizeable $\Delta a_{\mu}$ comparable to the potential anomaly. This is difficult to do within the MSSM. But if we have a much smaller $\Delta a_{\mu}$, absolute stability is also readily achieved within the MSSM, and BM2 is an example of this. For BM3, the DM direct detection cross section $\sigma_{\rm SI} \sim 3 \times 10^{-53} {\rm cm}^2$ is well below the sensitivity of the planned experiments.

BM4 illustrate that the non-minimal model can readily result in {\it both} a sizeable $\Delta a_{\mu}$ comparable to the potential anomaly and observable DM direct detection,  $\sigma_{SI} \simeq 10^{-47} {\rm cm}^2$, difficult to achieve in the MSSM. Furthermore, BM4 illustrates that the non-minimal model can readily have a stau NLSP, so that all selectrons/smuons decay into hard electrons/muons, enhancing observability. This is also difficult to achieve in the MSSM.

BM5 is realized in the non-minimal model, with slepton LSP, and 
$\Delta a_{\mu}$ comparable to the potential anomaly. We consider this spectrum to be realized in conjunction with $R$-parity violating couplings, so that the slepton LSP decays, either promptly or as a long-lived particle.

\subsection{MSSM Benchmarks}

 Two constraints from the observed spectrum of ordinary SM particles come from fitting the $m_{h^0} = 125$ GeV and the EW scale as represented by $m_Z = 91$ GeV. 
 For the reasons mentioned at the beginning of  \autoref{sec:OneLoopRGE}, we are approximating $A_{u,d}({\rm UV}) \approx 0$ at the KK matching scale, but including the running of $A$-terms below this. This results in modest IR stop-mixing,
 which translates into requiring multi-TeV stop masses to fit the observed Higgs mass. We have included this mixing effect. We have further validated the Higgs mass  for our benchmark points using FeynHiggs 2.18.1 \cite{Heinemeyer:1998yj,Heinemeyer:1998np,Degrassi:2002fi,Frank:2006yh,Hahn:2013ria,Bahl:2016brp, Bahl:2018qog}, requiring them to be consistent with the observed value given the estimated theoretical uncertainty of $\sim 1$ GeV for these calculations.   A second constraint is the requirement that EWSB takes place far below the $\sim 10$ TeV scale that typifies the individual $H_u$ IR soft terms, in particular $\mu$ and $M_{H_u}^2$. For  low-scale EWSB this requires these soft-terms to be fine-tuned to balance each other, 
\beq
\mu^2\simeq -m_{H_u}^2+\frac{m^2_{H_d}-m^2_{H_u}}{\tan^2\beta}
\eeq
thereby effectively removing $\mu$ from the list of independent variables.

\begin{table}[h]
\centering
\begin{tabular}{|c|c|c|}
 \hline
  Benchmarks & BM1 & BM2 \\ \hline\hline
 $\Delta a_{\mu }\times 10^{9}$ & 2.2 & 0.1 \\ \hline
 $m_{\tilde{e}_L}\approx m_{\tilde{\mu}_L}$ &0.54 & 0.75 \\
$m_{\tilde{e}_R}\approx m_{\tilde{\mu}_R}$ & 0.21 & 0.70 \\
 $m_{\tilde \tau_1}$ & 0.32 & 0.31\\
 $m_{\tilde \tau_2}$ & 0.63 &0.61\\ \hline
 $M_1$ & 0.20  & 0.30 \\
 $M_2$ & 0.93 & 1.4 \\
 $M_3$ &14 & 13\\
 $m_{\tilde t_{1,2}}$ & 10, 11 & 8, 9 \\
 \text{tan $\beta $} & 6 & 8 \\
 $\mu$  & 10.5 & 1.4 \\
 $m^2{}_{\text{Hd}}$ &-55 &110\\
 $\xi$ &0.086 & -0.095 \\\hline
\end{tabular}
\caption{Benchmark model parameters for MSSM. All quantities of mass dimensions are in units of $\TeV^{(2)}$. The first benchmark (BM1) accounts for the current muon $g-2$ anomaly, and has a metastable standard EWSB vacuum. The second benchmark (BM2) has an absolutely stable standard EWSB vacuum and shows the possibility of detectability in future DM direct detection  experiments, but with much smaller BSM contribution to muon $g-2$.
\label{tab:bmminimal}
}
\end{table}

In \autoref{tab:bmminimal} we present two realistic benchmark for MSSM, BM1 and BM2.
For both BM1 and BM2, sleptons of the first two generations are essentially unmixed after EWSB, with the gauge eigenstates being the mass eigenstates, and they are degenerate across these generations because their masses are determined by gauge couplings. On the other hand, the staus are more mixed after EWSB, and with significant contributions to their masses arising from the larger $y_{\tau}$ and $m^2_{H_d}$, they are heavier than other sleptons, 
an effect which is
calculable within the UV model.
Again, all the sleptons get significant mass contributions from their hypercharge $D$-term couplings to the Higgs scalars as well as from the two loop RG effects discussed in \autoref{sec:twoloopRGE}. 
As can be seen, we have taken $m_{H_{d(u)}}^2({\rm UV})$ to be large and {\it negative}, because this results in positive contributions to the stau masses. This in turn helps to solve the concerns raised earlier: (1) It
ensures that after EWSB-induced mixing the lightest $\tilde{\tau}_1$ mass eigenstate is not too light or tachyonic. (2) The trilinear stau-Higgs coupling can  (and does in our benchmarks) introduce a deeper EM-breaking ground state, but the stau masses in this benchmark are large enough to make the preferred EWSB ground state stable over cosmological timescales, as discussed in \autoref{sec:pheno}.

In the IR, $m_{H_u}^2({\rm IR})$ is corrected by top/stop loops to become even more negative, and this is fine-tuned against $\mu^2$ to result in the much smaller scale of EWSB via $H_u$ condensation. However, $m_{H_d}^2$ runs far less and therefore, with the positive $\mu^2$ contributions, $H_d$ remains non-tachyonic and very heavy. It therefore acquires only a small EWSB VEV ($\tan \beta \gg 1$) due to modest $B \mu$ Higgs-mixing.

We have large stop masses 
of $10$ and $11$ TeV for BM1 and $8$ and $9$ TeV for BM2, 
to give the large corrections to the $H_u$ quartic self-coupling needed to fit the $125$ GeV Higgs mass. These IR stop masses are generated predominantly by gluino-mediation, which requires $M_3 = 14$ TeV and $M_3 = 13$ TeV respectively in the IR. We have taken the bino to be the LSP, with mass only slightly smaller than the smuon/stau NLSP for BM1/BM2, so as to make the bino a viable co-annihilation Dark Matter candidate. 
Standard ${\rm \tilde{g}}$MSB models give $\tilde{\ell}_L$ mass contributions close to the wino mass, but in our case these are significantly cancelled against the $D$-term contribution as well as the two-loop effects, enhanced by the large squark and Higgs masses. 
 
Now we can focus on the differences between BM1 and BM2. BM1 matches the $g-2$ anomaly but, due to the small mixing of Bino with the heavy Higgsinos, the Bino is hidden from current and future dark matter direct detection experiments. 
Instead, in BM2, we take  the EW fine-tuning mainly  between $M_3$ and $M_{H_u}^2$, and have a lower Higgsino mass $\mu$. The direct detection rate is mediated by the Higgsino component of the bino-like LSP DM through the $t$-channel diagram, enabling a future detection possibility at experiments such as LZ/XENONnT~\cite{LUX-ZEPLIN:2018poe,XENON:2020kmp}. As discussed in \autoref{sec:nonminimalmodel}, in the MSSM this observability comes at the ``cost'' of having a very small muon $g-2$ correction. This is in contrast to our non-minimal model, to which we now turn.

\subsection{Non-minimal Model Benchmarks}

\begin{table}[h]
\centering
\begin{tabular}{|c|c|c|}
 \hline
  Benchmarks &  BM3: Absolutely stable EWSB vacuum & BM4: stau coannihilation  \\ \hline \hline
 $\Delta a_{\mu }\times 10^{9}$ & 2.0 & 1.8 \\ \hline
 $m_{\tilde{e}_L}$ & 0.40 & 0.34 \\
 $m_{\tilde{e}_R}$ & 0.34 & 0.34 \\
 $m_{\tilde{\mu}_1}$  & 0.36 & 0.33\\
 $m_{\tilde{\mu}_2}$  & 0.42  & 0.35 \\
 $m_{\tilde \tau_1}$  & 1.4 & 0.24 \\
 $m_{\tilde \tau_2}$  & 2.0 & 0.45 \\ \hline
 $M_1$   & 0.34 & 0.23 \\
 $M_2$  & 1.2 & 1.7 \\
 $M_3$ &8 & 8  \\
 $m_{{\tilde t_{1}},{\tilde t_{2}}}$ & 7 & (5, 6)  \\
 \text{tan $\beta $}  & 35  & 19 \\
 $\mu$   & 2.5 & 2.4 \\
 $\mu_u$  & 7.7 & 5   \\
 $\mu_d$   & 8  & 2.2 \\
 $\mu'$  & 1 & 10   \\
 $m^2_{{H_d}}({\rm UV})$  &-55 &-1.4  \\
 $m^2_{{H^\prime_d}}({\rm UV})$  &430 & 900 \\
 $\xi$ & - 0.05 & -0.05  \\
 $\xi^\prime$  & -0.025 & -0.019 \\
  $m^2_{\text{colored} \, H'}({\rm UV})$  &400 & 900\\  \hline
\end{tabular}
\caption{ Benchmark model parameters for the non-minimal model. All quantities of mass dimensions are in units of $\TeV^{(2)}$.  In the first benchmark (BM3), the potential has no deeper minimum and the standard EM-preserving EWSB vacuum is stable. The second benchmark (BM4) shows the possibility for bino DM coannihilating with stau. It has a long lived, but not absolutely stable, standard vacuum.
BM4 also features detectability from future direct detection experiments as well as a sizeable muon $g-2$ correction, enabling multiple and consistent discoveries across disparate probes. 
\label{tab:bmnonminimal}
}
\end{table}
In \autoref{tab:bmnonminimal}, we present two benchmarks for the non-minimal model which illustrate two qualitatively interesting possibilities that are difficult to achieve in the MSSM. We have explored the scalar potential of the 
benchmark in the first column (BM3) and found that it has an absolutely stable standard EWSB vacuum, so that there is no reliance on cosmic history to put us in a metastable vacuum or the need to compute its lifetime. The staus have gained a significant increase in  mass compared to the other sleptons, from $y_\tau$-mediation due to relatively large and negative $m^2_{H_d}$ and large $\tan \beta$. Still the non-minimal structure of the model allows for keeping $\mu \approx 2.5$ TeV, small enough so that the trilinear coupling does not lead to a deeper EM breaking vacuum. The smuon and bino masses, with the enhancement from $\mu$ and $\tan \beta$, can account for the muon $g-2$ anomaly. $\mu_u$ is chosen so that it fine-tunes EWSB to be small (relative to the $\sim 10$ TeV scale), that is,
\beq
|\mu_u|^2\approx -m^2_{H_u}({\rm IR})-|\mu|^2.
\eeq
Large $H_d$ condensation is avoided by choosing large enough $\mu_d= 8$ TeV$ (\mu_d^2>-m^2_{H_d})$. Also $m^2_{{H'_u}}\approx m^2_{{H'_d}}$ are large enough to prevent a large VEV for $H^\prime_u$ (eq.~\eqref{eq:Huprimevev}). 

The benchmark in the second column (BM4) shows a point with the lighter stau, $\tilde{\tau}_1$, as the NLSP, and with bino slightly lighter, allowing for coannihilating bino dark matter. Having stau as the NLSP allows for the other sleptons to be significantly heavier. As we will discuss in the next section, this has an important phenomenological consequence, allowing for a more $e/\mu$-rich signature at the LHC.  
With the lighter stau masses in this benchmark, the standard EWSB vacuum is not absolutely stable, but still $\mu$ has been chosen small enough for having a meta-stable standard vacuum over cosmological time scales. For this point, the value of $m^2_{H_u}$ in the UV is not very large and its IR value is dominantly due to one-loop RG effect via top/stop loop. 
Also different from our other benchmarks, $m^2_{H_d}$ is not large, and therefore the slepton mass squareds are approximately universal. It is then the significant mixing between the singlet and doublet staus that makes one of the mass eigenstates lighter, the NLSP, and the other one heavier than the other sleptons. The choices for other parameters follows along the lines described for BM3 benchmark.
Further, the lightest Higgsino in BM4 is at 1.1 TeV, with predominantly $\tilde H_{u,d}$ components, allowing BM4 to be accessible in dark matter direct detection experiments, along the lines discussed in \autoref{sec:nonminimalmodel}. 

Finally in both benchmarks, we have also shown the choice for the colored unification partners of $H^\prime_{u,d}$. To preserve gauge coupling unification, their masses should not be too far from the uncolored $H^\prime_{u,d}$. For simplicity, we have taken the soft masses for the two color-triplet $H^\prime$ to be equal, although that is not necessary. These large soft masses also feed into the slepton masses via two loop RG effects, similar to the effects discussed in \autoref{sec:twoloopRGE} for the MSSM. We have included all such effects for the benchmarks presented here.

\subsection{$R$-Parity Violating Benchmark in the Non-minimal MSSM}

It is also  plausible to have R-Parity Violation (RPV), for reviews, see e.g.,~\cite{Barbier:2004ez}. The considerations in terms of model parameters will change slightly but significant changes in phenomenology will occur. 
Note that we are not using RPV couplings to mediate muon $g-2$ contributions, but rather just to explore alternate LHC phenomenology, so that our RPV couplings can be quite small. See ref. \cite{Altmannshofer:2020axr, BhupalDev:2021ipu} for recent work on substantial RPV couplings which directly contribute to the muon $g-2$ (and other flavor-dependent) observables.

\begin{table}[h]
\begin{tabular}{|c|c|c|c|c|c|c|c|c|c|}
 \hline
 $\Delta a_{\mu }\times 10^{9}$ &  $m_{\tilde{e}_R}\approx m_{\tilde{\mu}_1}$ &   $m_{\tilde{e}_L}$ &   $m_{\tilde{\mu}_2}$ & $m_{\tilde{\tau}_1}$ & $m_{\tilde{ \tau}_2}$ & $M_1$ & $M_2$ & $M_3$ & $m_{\tilde{t}_1}$, $m_{\tilde{t}_2}$ \cr \hline
 1.7 & 0.47 & 0.50  & 0.52 & 1.4 & 1.9 & 0.60 & 1.3 & 8 & 6, 7 \cr \hline  \hline 
 $m^2_{{H_d}}({\rm UV})$ & $m^2_{{H^\prime_d}}({\rm UV})$ & \text{tan $\beta $} & $\mu$ &  $\mu_u$ &  $\mu_d$ &  $\mu^\prime$ &  $\xi$ & $\xi^\prime$ &  $m^2_{\text{colored}\, H^\prime}({\rm UV})$ \cr \hline
 -50 & 580 & 35 & 6.1 & 6.5 & 8 & 1  & -0.02 & -0.012 & 400 \cr \hline
\end{tabular}%
\caption{Benchmark model (BM5) with slepton LSP and $R$-parity violation. All quantities with mass dimensions are in units of $\TeV^{(2)}$. The standard EWSB is meta-stable, with lifetime larger than the age of the universe. 
\label{tab:bmminimalRPV} 
}
\end{table}

In \autoref{tab:bmminimalRPV}, we present a benchmark (BM5) with a charged slepton LSP. The singlet selectron and the lighter smuon mass eigenstate are almost degenerate in mass. The stau masses are dominated by the $y_\tau$ mediated RG contribution, and the similar contribution to the smuon mass squared mediated by muon Yukawa coupling is also non-negligible in this benchmark due the large $\tan \beta$ and $m^2_{H_d}$. With the heavier slepton spectrum in this benchmark, larger $\tan \beta$ (and larger $\mu$ compared to the benchmarks of \autoref{tab:bmnonminimal}) have been chosen to account for the current muon $g-2$ anomaly.

\section{LHC Phenomenology}
\label{sec:pheno}

Having understood the model construction and generic parameter space, and laid out the benchmarks in the previous section in \autoref{tab:bmminimal} and \autoref{tab:bmnonminimal}, we describe the corresponding collider tests. We first describe the $R$-parity conserving phenomenology with a bino LSP. Then we discuss the rich phenomenology associated with $R$-parity Violation. We focus on the current constraints and the critical future searches at the (HL-)LHC.

\subsection{$R$-Parity Conserving Phenomenology}

First, we discuss the minimal signatures from {\it just} the bino LSP and slepton NLSP, the central players in co-annihilating dark matter. Obtaining the thermal relic abundance of bino dark matter in the early universe requires that these two states are nearly degenerate, to within 
3-8 GeV 
for our benchmark models. 
In the benchmark models, BM1, BM2~(\autoref{tab:bmminimal}), BM3 and BM4~(\autoref{tab:bmnonminimal}), the nearly degenerate bino LSP and slepton NLSP~\footnote{These are smuon NLSP for BM1 and BM3, and stau NLSP for BM2.} will provide minimal missing energy signatures, which can be continuously improved at the LHC but is limited by the large background and systematic uncertainties.

Beyond the minimal missing energy LSP, NLSP searches, our models also feature other sleptons (heavier selectron, smuon, and staus, as well as sneutrinos) and winos, ranging from the weak scale to TeV scale. Given the importance and kinematic accessibility of the sleptons in our model, we emphasize it is literally a ``sleptonic SUSY'' phenomenology as far as the LHC is concerned. Compared to the more compressed NLSPs, the heavy slepton pair production and decays to LSP will provide more energetic leptons in the final states. 

The phenomenology here divides into sleptons decaying into 
somewhat soft (but 
still
visible)-leptons, and more non-compressed sleptons decaying into hard leptons. 
 For BM3 (BM4), one can look for (both the lighter and) the heavier eigenstates of the smuons and selectrons, ($\tilde \mu_1$ and $\tilde e_1$ as well as) $\tilde \mu_2$ and $\tilde e_2$, which have about 50-100 GeV mass splitting with the LSP, giving rise to 
softish
($p_T<100$~GeV)-leptonic SUSY phenomenology. 
The searches for pair-produced $\tilde \mu^\pm_2\rightarrow \tilde \chi^0_1+\mu^\pm$ and $\tilde e^\pm_2\rightarrow \tilde \chi^0_1+e^\pm$ call for improvement in current LHC SUSY searches of 
softish
leptons plus missing energy~\cite{Han:2013gba,Han:2014nba,ATLAS:2019lff,ATLAS:2019lng,CMS-PAS-SUS-18-004}. 
Despite the challenging nature of this channel from the trigger, signal rate, and SM background, given the multiple motivations for Sleptonic SUSY the current searches should be improved as much as possible by exploring the soft-lepton frontier. The experimental program at the LHC has started to probe such soft signatures, typically with hard initial state radiation (ISR)~\cite{ATLAS:2019lng,CMS-PAS-SUS-18-004}. Further upgrades and development in the experimental search program could provide definitive information about the allowed model parameter space. 

Further, there are interesting opportunities to search for heavier sleptons with larger mass splitting with the LSP, e.g., for (heavier) selectron and smuon in BM1 and BM2, and the (heavier) staus in BM1, BM2, BM3 and BM4. One can improve the selectron and smuon searches in the non-compressed regime with higher statistics from the HL-LHC. (S)tau decays and their tagging suffer lower efficiency and more background. Consequently, if the stau is not too heavy, such as in BM1 and BM2, the searches for stau plus missing energy, potentially with ISR, could help probe the model further~\cite{CMS:2019zmn}. 
The charged-current production of charged sleptons plus sneutrinos has a larger rate than neutral-current slepton production. However, the sneutrinos decay invisibly to LSP plus neutrinos in this scenario.  Therefore, while the charged current typically provides a large production rate, the searches might suffer a larger background and, hence, be less promising than the neutral current production of charged sleptons~\cite{Alvestad:2021sje}. Nevertheless, new searches could be designed to capitalize on the charged current production.

Amongst these new (sub)TeV states, wino pair-production through neutral- and charged-current processes provide distinctive signatures. The wino phenomenology is somewhat different from conventional SUSY discussions. Due to the scale separation between the wino soft mass $M_2$ and large Higgsino mass $\mu$, the dominant decays of the winos are through sleptons, not directly into electroweak gauge bosons plus the LSP bino. For the wino-chargino $\tilde \chi_1^\pm$, about half of the decays provide a visible charged lepton plus missing energy; the rest decay into neutrinos plus compressed sleptons. For wino-neutralino $\tilde \chi_2^0$, the decays into sneutrino and neutrino pairs are invisible. The visible decays include the decays into heavy~(light) smuon/selectron and muon/electron pairs that provide a hard muon/electron and a soft muon/electron and missing energy, and the decays into staus and taus that provide two visible taus and missing energy. Overall, the wino pair production has a sizable rate and can give rise to multi-lepton (3 or 4 leptons) signatures, which is  very promising  to look for at the LHC~\cite{ATLAS:2019wgx,CMS:2021cox,CMS:2022nty}. In fact, these multilepton searches provide one of the leading constraints on our model, that pushes the wino mass to be beyond 900~GeV for our benchmarks. 
In particular, our signals typically yield mixed  lepton flavors, off-$Z$ multi-leptons, that would benefit from the improvement in these ongoing searches at the LHC.

\subsection{$R$-Parity Violating Phenomenology}

It is entirely plausible that $R$-parity is violated (RPV) so that all superpartners are unstable, and dark matter has its origins in some other sector. This gives rise to new and interesting experimental opportunities. The LSP could be the light smuon~\footnote{The selectron can be a few GeV heavier and its major decay will be through the same RPV coupling (assuming it is lepton-flavor blind). So the phenomenology of smuon also applies to selectrons.}, such as in BM5 shown in \autoref{tab:bmminimalRPV}. In this case, the phenomenology becomes very rich, as one is required to turn on a tiny RPV coupling to make the light smuon unstable so as to be compatible with cosmology. For instance, one can turn on the $L Q d^c$ superpotential operator that facilitates decays of the LSP smuon decays into dijets. So long as one does not simultaneously turn on baryonic RPV couplings that introduce severe constraints from proton decay, the $L Q d^c$ RPV coupling $\lambda^{\prime}$ can be as large as $10^{-3}$~\cite{Barger:1989rk,Allanach:1999ic,Chamoun:2020aft} without this new source of flavor violation being constrained by flavor tests. Small R-parity symmetry-violation is of course a natural possibility.

As long as $\lambda^{\prime}>10^{-7}$, the light sleptons decay promptly, so there is a sizeable window where this happens and where there are no experimental flavor constraints. The simplest direct search for pair-produced smuons give rise to the signatures of dijet resonance pairs, and the limits are relatively weak at the LHC~\cite{ATLAS:2017jnp,CMS:2018mts} due to the large QCD background, around 120~GeV (from our estimation and reinterpreation of these searches). Instead, one should look for the more interesting phenomenology of pair production of the sneutrinos and heavy smuons. This new channel will generate promising signatures of charged leptons from the intermediate $Z$ or $W$ boson decay, plus a pair of dijet resonances. Even better, the pair-produced stau states will give rise to 4 leptons plus dijet resonance pairs. Searches for these lepton-rich dijet pair channels are promising at the LHC~\cite{Brust:2011tb} and could potentially cover the allowed parameter space of our model. 

For $\lambda^{\prime}<10^{-7}$, the slepton LSP is long-lived, producing displaced decays or even heavy (collider-)stable charged particle (HSCP) tracks at the LHC. The searches for displaced vertices and HSCP have gained a lot of attention in the current LHC program, and will improve significantly in the near future. The displaced vertex search rules out weakly produced particles below $\sim900$~GeV~\cite{Liu:2015bma,Liu:2018wte,CMS:2020iwv}, leaving room only for a longer lifetime of the smuon LSP. Current HSCP search rules out sleptons below $\sim460$~GeV~\cite{Khachatryan:2016sfv,Aaboud:2019trc}, leaving BM5 of our model viable, with a RPV coupling $\lambda^{\prime}<10^{-9}$. Instead of looking for the smuon LSP in future searches, the HSCP will be produced in a prompt lepton-rich fashion through the decays of heavier electroweakinos and heavier sleptons. The searches for HSCP plus leptons provide new LHC opportunities that provide a promising venue to test our model. Beyond the search for discovery from this channel, the long-lived signatures will have low background and no ambiguity from the combinatorics to reconstruct the underlying new state. Hence, one can immediately access the quantum numbers of the newly discovered states from this channel through the production rate, decay location distribution, angular distribution, lepton flavor tagging, etc. One can then identify the SUSY nature of the underlying physics readily.

\section{Conclusions}
\label{sec:conclusions}

We have argued in this paper that Sleptonic SUSY is an attractive framework for new physics, from a variety of angles. 
Most broadly, it is truly remarkable that the SUSY paradigm for particle physics  can be realized in a  relatively elegant and economical manner in terms higher-dimensional sequestering. This harmoniously results in  gaugino-mediated, $D$-mediated and Higgs-mediated SUSY breaking, in concert with the Giudice-Masiero mechanism for a comparable $\mu$ parameter.  These SUSY-breaking mechanisms automatically respect and generalize the standard GIM mechanism, and robustly can avoid new CP-violating phases, thereby satisfying the tightest flavor and EDM bounds that tightly constrain {\it any} paradigm beyond the standard model. The higher-dimensional framework can also be simply extended to realize orbifold grand unification, with precision gauge-coupling, $b-\tau$ unification and doublet-triplet splitting. Therefore, it is well worth asking what more specific incarnations of this structure are motivated by our current experimental situation. 

The absence of colored superpartners thus far in LHC searches and the heaviness of the $125$ GeV Higgs scalar suggest stops at $\sim 10$ TeV or beyond, but this still leaves an opening for other superpartners to be within LHC reach. In this way, Sleptonic SUSY, with its sub-TeV uncolored superpartners and colored and $\sim 10$ TeV Higgs superpartners, stands out, as both easily realizable within the above theoretical framework and phenomenologically accessible. Additionally, while some options for realizing the original hope that the LHC can re-create thermal-relic Dark Matter particles are now ruled out, Sleptonic SUSY readily incorporates viable co-annihilating dark matter at the bottom of its spectrum, 
which is potentially discoverable at the LHC.
It is of course very intriguing that there exists a potential anomaly between the muon magnetic moment measurement and standard model theory, and that Sleptonic SUSY again stands out as one of the most plausible ways of explaining this consistent with an extended GIM mechanism.\footnote{While a sizeable smuon contribution to the muon $g-2$ is a robust consequence of Sleptonic SUSY,  its magnitude is quite sensitive to the details of the model parameters. This can be consistent with the nominal size of the potential anomaly, $\Delta a_{\mu}= 259\pm 60 \times 10^{-11}$. But if the experimental anomaly or theoretical computation change in the future, such changes can still readily be accommodated by modest changes within the Sleptonic SUSY spectrum.}

The Little Hierarchy Problem, of why new physics has not already appeared at the weak scale to ensure its radiative stability, is an outstanding puzzle in beyond-standard model theory in general. The Anthropic Principle operating within a large multiverse with varying effective field theories has been offered as an alternative paradigm to the naturalness principle underlying the hierarchy problem. But it is quite possible that neither principle is outright dominant, but rather anthropic selection ``frustrates'' naturalness, leaving a little hierarchy problem. While a detailed understanding of anthropic selection and multiverse distributions is still lacking, a 
 plausible corollary of such selection is the 
 Principle of Living Dangerously in which the IR dynamics should be close to an anthropically-significant phase transition. From this viewpoint, it is intriguing that Sleptonic SUSY is locally
  trapped between two ``dangerous'' phase transitions: if we de-tune the theory to make it slightly more natural, in one direction we pass to a phase of unbroken electroweak symmetry while in the other direction we pass to a phase in which the photon is massive! 

We have generalized UV models of Sleptonic SUSY in a variety of ways, more fully covering the MSSM parameter space, especially the regime of moderate $\tan \beta$ as well as the RPV scenario and phenomenology. Furthermore, we developed and studied a non-minimal model in which the cosmological stability constraints of the electroweak vacuum are mitigated, thereby allowing richer phenomenological and cosmological options. 
For simplicity, we considered the regime in which UV $A$-terms are subdominant. While we argued they are not qualitatively important, it would be useful to include them in a more thorough exploration of the phenomenology, in particular that of the  stops.

Experimentally, the general Sleptonic SUSY scenario has many interesting  facets relevant to  discovery  at the LHC and future colliders. Overall, sleptons and electroweakinos at weak-TeV scales call for more phenomenological and experimental explorations. In the bino DM and slepton NLSP case, while continuing to improve missing energy searches at the LHC, one can also focus more on the associated heavier sleptons that provide more visible leptons in the semi-compressed regime. Further, in the RPV case, one can conduct new searches that significantly enhance the discovery potential of current searches by requiring additional prompt leptons in the final state, for both promptly-decaying LSP searches and long-lived LSP searches. These new search channels could well be the simultaneous discovery channels for LSP and heavier states. If the sleptonic SUSY spectrum is highly compressed, rendering these channels very challenging at the LHC, the LHC can still look for the $\sim$ TeV scale Winos decaying into the various weak scale sleptonic states, whose reach will improve steadily with higher luminosity. 

Beyond discovery, one can more ambitiously try to pin down 
quantum numbers of the new states to provide evidence that supersymmetry is at work, e.g. to distinguish from other possible BSM scenarios 
such as
fermionic ``lepton partners'' or other weakly-charged new states. In the bino DM and slepton NLSP case, one needs sizable statistics to extract the angular correlations from backgrounds, which is not easy to achieve even at HL-LHC given the current exclusions. Interestingly, in the long-lived RPV scenario, e.g., BM4 with small RPV coupling, one can pair-produce  binos, which provide prompt leptons to trigger on and the low-background displaced-vertices can extract the spin-quantum number more easily, providing an exciting program of LHC studies.

Even if a modestly compressed Sleptonic SUSY spectrum evades discovery at the LHC, it would be guaranteed discovery at future colliders. Future lepton colliders, such as ILC, CLIC or muon colliders, can readily achieve center-of-mass energy greater than twice the $\sim$ weak-scale slepton mass. We will then be able to make definitive discoveries through slepton pair production in all scenarios, and determine the quantum numbers through production rates, differential observables, etc. 
Even at the low-energy stage of future lepton colliders, operating at the $Z$-pole, it may be possible to detect the shift in the $Z \overline{\tau} \tau$ coupling loop-mediated by weak-scale staus, due to the large mixing that connects to the large shifts in $g-2$ (of muon and $\tau$), the near-instability of the EW vacuum, and frustrated naturalness. Staus this light, consistent with cosmologically stable EW vacuum, are most straightforwardly achieved within our non-minimal model. 

In this paper, we have focused on the limit in which the GIM mechanism is perfectly satisfied, with negligible new sources of BSM flavor violation. However, it is possible that there are observably large deviations from such a limit.  This can obviously happen if there are sizeable RPV couplings. But even with R-parity, as we reviewed, extra-dimensional sequestering can be imperfect, introducing small flavor-violating SUSY-breaking effects. Either case can lead to opportunities for new signals in flavor physics, in particular associated to the third generation.

At future hadron colliders, such as FCC-hh and SPPC, we will be able to make a full discovery by observing the 10~TeV scale squarks and gluinos. Even at a lower center-of-mass energy below the production of the colored states, we will likely make the first discovery by observing the Winos and their decays into  lepton-rich final states. 

We have seen that Sleptonic SUSY, especially in the regime with sizeable corrections to the muon magnetic moment, can readily have a deeper vacuum than the standard EW vacuum, in which staus condense and Higgs electromagnetism. While the EW vacuum can have cosmological lifetime at low temperatures, it will be important to study a fuller cosmological history to 
assess whether the EW vacuum is stable enough when thermal effects are included starting from inflationary reheating.
It will also be important to improve upon the vacuum lifetime estimates presented here. It is however possible to find regions of parameter space in which the EW vacuum is absolutely stable, in particular in  our non-minimal model.

\acknowledgments 
We would like to thank Z. Chacko, Junwu Huang, Keith Olive, Riccardo Rattazzi, and Tuhin Roy for helpful discussions.
We would also like to thank Tsutomu Yanagida for bringing ref. \cite{Inoue:1991rk} to our attention.
KA, ME and RS were supported in part by the NSF grant PHY-1914731 and by the Maryland Center for Fundamental Physics. ME was also supported by the Swiss National Science Foundation under contract 200020-188671 and through the National Center of Competence in Research SwissMAP.
ZL was supported in part by the U.S. Department of Energy (DOE) under grant No. DE-SC0022345. ZL would like to thank the Aspen Center for Physics, supported by National Science Foundation (NSF) grant PHY-1607611, where part of this work was completed.

\newpage

\appendix

\section{A Toy Model for Vacuum (Meta-)stability Considerations}

This appendix is not necessary for the numerical estimates of the EW metastable vacuum lifetime, which we performed as outlined in the main text, but it does provide a useful analytic toy model with only one scalar field which makes the essential/parametric considerations of the vacuum stability analysis involving $H_u$ and stau fields transparent.
Consider a single real scalar field $\phi$ with the following Lagrangian:
\beq
\mathcal{L}=\frac{1}{2} \left(\partial \phi \right)^2-\frac{1}{2}m^2 \phi^2+\frac{1}{3} \delta \, \phi^3-\frac{1}{4} \lambda \, \phi^4.
\eeq
To connect with the potential of eq.~\eqref{eq:thepotential}, we can think of $\phi$ as the distance in the field space from the standard EM preserving 
vacuum along the direction connecting the standard vacuum to the possible EM breaking vacuum.
 $m^2$ then corresponds to a combination of the stau and the SM Higgs mass terms, which if stau masses are larger, is dominantly set by $m^2_{L_3}$ and $m^2_{\bar{e}_3}$. The cubic coupling $\delta$ corresponds parametrically to $y_\tau \mu$ and $\lambda$ to a combination of quartic couplings set by the gauge couplings, $y_\tau$, and the one loop Higgs quartic $\delta \lambda_H$ induced by top and stop loops.

The potential for this field (assuming $m^2>0$) has a minimum at $\phi=0$ (corresponding to the standard vacuum). The condition for this to be the absolute minimum of the potential is same as the condition for the potential being non-negative for all $\phi$ which is 
\beq \label{eq:absolutestabilitycondition}
\delta^2<\frac{9}{2} m^2 \lambda.
\eeq
This corresponds parametrically to smallness of the ratio $\mu y_\tau/( g \, m_{\tilde{\tau}})$.

Now let us assume that this condition is not satisfied and there is another deeper minimum for the potential. We can then look for the action of the bounce solution for tunneling out of the $\phi=0$ vacuum. By a field and coordinate rescaling, the Euclidean action for the $O(4)$-symmetric bounce can be put in the following form:
\beq 
S=2\pi^2 \frac{m^2}{\delta^2} \int d\tilde{r} \tilde{r}^3 \left[\frac{1}{2} \left( \frac{d \tilde{\phi}}{d \tilde{r}} \right)^2 +\frac{1}{2}\tilde{\phi}^2-\frac{1}{3} \tilde{\phi}^3+\frac{1}{4} \frac{\lambda m^2}{\delta^2} \tilde{\phi}^4 \right].
\eeq
The bounce is a solution to the equation of motion following from the action above with the boundary conditions $\tilde{\phi} \rightarrow 0$ as $\tilde{r} \rightarrow \infty$ and $\frac{d\tilde{\phi}}{d\tilde{r}}=0$ at $\tilde{r}=0$ \cite{Coleman:1977py}. Since the boundary conditions do not introduce any further dependence on the parameters of the potential, the bounce action has the following parametric form:
\beq
\label{eq:bounce}
S_B\approx 204 \frac{m^2}{\delta^2} f_B\left( \frac{\lambda m^2}{\delta^2} \right)
\eeq
The prefactor has been chosen such that $f_B(0)=1$ and it has been fixed by numerically solving the bounce for $\lambda=0$. For small $\frac{\lambda m^2}{\delta^2}$, $S_B$ can be perturbatively corrected by expanding in $\frac{\lambda m^2}{\delta^2}$. We obtain numerically, for small $x$,
\beq
\label{eq:f}
f_B(x)\approx 1+ 1.4 \frac{9 x}{2}+  1.9 \left( \frac{9 x}{2}\right)^2+ 2 \left( \frac{9 x}{2}\right)^3+ \mathcal{O} (x^4).
\eeq
Also we know from eq.~\eqref{eq:absolutestabilitycondition} that $\lim_{x\rightarrow \frac{2}{9}} f(x)=\infty$, i.e. the lifetime should become infinite as the $\phi\neq0$ minimum becomes degenerate with the $\phi=0$ one. 
In terms of the parameters of the potential of eq.~\eqref{eq:thepotential},
for large $\mu$ and thus large $\delta$, we can set $f \approx 1$ in eq.~\eqref{eq:f} so that $S_B \propto \frac{m^2_{\tilde{\tau}}}{(\mu \, y_\tau)^2}$ using eq.~\eqref{eq:bounce}.
As we decrease $\mu$ (keeping other parameters fixed), 
bounce action  increases: 
at some point, $\delta$ becomes small enough that the full expression for $f$ in eq.~\eqref{eq:f}
has to be used. 
The bounce action continues to increase as we reduce $\mu$ further 
until 
 as $\mu$ decreases, 
it diverges when the EM conserving and the EM breaking vacua become degenerate.
In that limit the bounce action can be computed analytically using the thin-wall approximation giving
\beq
S_B \approx \frac{3 \pi^2 m^2}{4 \delta^2 \left(1-\frac{9 \lambda m^2}{2 \delta^2} \right)^3}\approx 7.4 \frac{m^2}{ \delta^2 \left(1-\frac{9 \lambda m^2}{2 \delta^2} \right)^3}.
\eeq

In the models considered here, unless $\tan \beta$ is very large, $\lambda$ is set primarily by the gauge couplings and is not sensitive to the other parameters. Then to maximize the lifetime of the vacuum (or to achieve absolute stability), one can maximize the ratio $\frac{m^2}{\delta^2} \sim \frac{m^2_{\tilde{\tau}}}{(\mu \, y_\tau)^2}$. In the MSSM realization of Sleptonic SUSY, 
without the help from $y_\tau$-mediated contribution to the stau soft mass-squareds proportional to large and negative $m^2_{H_d}$, eq.~\eqref{eq:RGEHiggsforL3E3}, this ratio is too small to lead to a cosmologically meta-stable vacuum, once the $125$ GeV Higgs, EWSB scale and the muon $g-2$ are fit,  and it is essential to use the yukawa/Higgs mediation RG effects to increase the stau masses. Moreover, importantly, increasing this effect does not indefinitely increase the vacuum lifetime, and is not enough to remove the EM breaking minimum. To see this assume that $|m^2_{H_d}|$ and/or $\tan \beta$ is large enough so that the stau masses are dominated by the $y_\tau$-mediation effect, that is $m^2_{\tilde{\tau}}\propto - m^2_{H_d} \tan^2 \beta$. In this regime, $\mu$ also needs to be large enough ($|\mu|^2+m^2_{H_d}>0$) to prevent large $H_d$ condensation. Therefore, we see that in this regime the ratio  $\frac{m^2}{\delta^2}$ is at most of order one, and it turns out that it is numerically not large enough to lead to an absolutely stable vacuum. In fact in this limit, the bounce action turns out to be  close to the limit for the vacuum to have a lifetime of the order of the age of the universe today.

However, in the non-minimal model, $\mu$ is not constrained to be 
large
by $m^2_{H_{d,u}}$ anymore, as the non-condensation of $H_d$ can be achieved using other generalized $\mu$-terms. 
Since $\mu$ 
can then be chosen to be smaller 
in the non-minimal model and 
$\delta \propto \mu$, we see that 
$\delta$, in turn, is allowed to be smaller, enough for having an absolutely stable standard EWSB vacuum.

\section{Other Charge- and/or Color- breaking Vacua} \label{app:othervacua}
As it was mentioned in \autoref{sec:stability}, there may be minima of the potential with nonzero VEVs for fields that are not included in the potential of eq.~\eqref{eq:thepotential}. In this appendix we briefly discuss the regions in  field space where such charge- and/or color-breaking vacua my lie. 

In the absence of SUSY-breaking, there exist paths in the field space along which the potential is (approximately) flat, that is the potential arising from the $D$-terms and $F$-terms vanish.
The soft SUSY breaking effects can either lift the potential along these paths, or lower it which can lead to new minima of the potential, or even possibly to potentials that are unbounded from below \cite{Komatsu:1988mt,Casas:1995pd,Strumia:1996pr}. 
For the models studied
%
%
in this paper, we have considered scenarios where the soft mass squared terms for $m^2_{H_{u,d}}$ are negative. So the potentially problematic flat paths involve the associated Higgs fields. In the following, we first consider the MSSM and then later briefly comment on the corrections in the non-minimal model. In particular the following three paths can lead to charge- and/or color- breaking vacua:

\textbf{(i)} $H_u,\tilde{\tau}_L,\tilde{\tau}_R, \tilde{\nu}_{i\neq 3} \neq 0$ and other fields set to zero. Only one of the sneutrinos of the first two generations need to be non-zero. The $F$-term for $H_d$ can be set to zero relating the staus and $H_u$, $\tilde{\tau}_L \tilde{\tau}_R^* =-\frac{y_\tau}{\mu} H_u$,
and the $D$-term potential vanishes if
\beq
|\tilde{\nu}|^2=|H_u|^2+|\tilde{\tau}_L|^2=|H_u|^2+|\tilde{\tau}_R|^2
\eeq

\textbf{(ii)} $H_u,\tilde{b}_L,\tilde{b}_R, \tilde{\nu} \neq 0$ and other fields set to zero, similar to path (i), with sbottoms instead of staus  cancelling the $F$-term for $H_u$. 
The $SU(3)$ $D$-term also vanishes if the two sbottom fields have equal magnitudes.

\textbf{(iii)} $H_d^0,\tilde{t}_L,\tilde{t}_R, \tilde{e}_L,\tilde{e}_R \neq 0$: 
and all other scalars set to zero.
Setting $|\tilde{t}_L|=|\tilde{t}_R|\equiv \tilde{t}$, the $SU(3)$ $D$-term vanishes.
Then setting $|\tilde{t}^2|=|\frac{\mu}{y_t} H_d^0|$, and choosing appropriate phases/signs, the $F$-term for $H^0_u$ vanishes. 
The charged component of the doublet selectron $\tilde{e}_L$ can be used to set the $SU(2)$ $D$-term to zero, and finally the hypercharge $D$-term can be set to zero using the $SU(2)$-singlet selectron,
\beq
|\tilde{e}_R|^2=|\tilde{e}_L|^2=|H_d^0|^2+|\tilde{t}^2|.
\eeq
The supersymmetric potential along this path is however not exactly flat since there are quartic couplings $\propto y_e^2$ from $F$-terms corresponding to $H_d$ and slectrons, but these quartic couplings are very small, $y_e^2\sim 10^{-11} (\tan \beta)^2$.

To see the possibility of the potential charge and/or color breaking vacua more clearly, one can write the potential in terms of only one of the nonzero fields, and including the soft SUSY breaking terms. For example for path (iii), in terms of $H_d^0\equiv \varphi$,
\beq \label{eq:potHdstop}
V(\varphi)=  (m^2_{H_d}+m^2_{L_1}+m^2_{\bar{e}_1})|\varphi|^2+\frac{|\mu|}{y_t} (m^2_{Q_3}+m^2_{\bar{u}_3}+m^2_{L_1}+m^2_{\bar{e}_1})|\varphi| +\delta \lambda_{H_d} |\varphi|^4
+\mathcal{O} (y_e^4),
\eeq
where $\delta \lambda_{H_d}$ is the one-loop  quartic coupling for $H_d$ dominantly induced by bottom/sbottom loops. This shows that if $m^2_{H_d}+m^2_{L_1}+m^2_{\bar{e}_1}<0$, the EWSB vacuum can be destabilized.

Having outlined the different directions in which deeper vacua than the EWSB vacuum can be encountered, we of course need  the potential energy to be bounded from below in order for the theory to be physically sensible. In the renormalizable limit of the MSSM, in which we also include Dirac neutrinos, the charged lepton Yukawa coupling effects (such as the last term in eq.~\eqref{eq:potHdstop} above) and neutrino Yukawa coupling effects give rise to (small) quartic couplings that bound the potential at very large field values. Further, there may be other stabilizing effects in the very far UV where the MSSM is completed into a unified or string theory. None of these effects would be relevant at the moderate field values which dominate the tunneling processes determining the EWSB vacuum lifetime, and are therefore dropped in the related estimates. 

In the non-minimal model, the $F$-terms for $H^\prime$ fields are nonzero along the above paths. In order to set those $F$-terms to zero,  one of the $H^\prime$ scalars should be non-zero, but then the large positive $m^2_{H^\prime}$ stabilize the potential. 
Also with vanishing $H^\prime$, the positive contribution of $H^\prime$ F-terms, lead to a more stable standard vacuum.
In particular, for path (iii), the $H^\prime_u$ F-term adds a contribution $|\mu_d|^2 |\varphi|^2$ to the potential in eq.~\eqref{eq:potHdstop}. Therefore a choice of $|\mu_d|^2>-m^2_{H_d}$ would be sufficient to avoid a minimum along this path.  
\newpage
\bibliographystyle{utphys}
\bibliography{refs}

\end{document}